\documentclass[a4paper,11pt]{article}
\usepackage{jheppub}

\usepackage[most]{tcolorbox}
\usepackage{blkarray, bigstrut} %
\usepackage[english]{babel}
\usepackage{amsmath}
\usepackage{microtype}

\usepackage{esint}
\usepackage{babel}
\usepackage{color}
\usepackage{enumitem}
\usepackage{subfigure}

\usepackage{float}
\usepackage[noabbrev]{cleveref}
\usepackage{bbold}
\usepackage{amsfonts}
\usepackage{amssymb}
\usepackage{amsthm}
\usepackage{enumitem}
\usepackage{shuffle}
\usepackage{comment}
\usepackage{cancel}
\usepackage{blkarray}
\usepackage{multirow}
\usepackage{subfigure}
\usepackage{longtable}
\usepackage{savesym}
\savesymbol{div}
\usepackage{physics}
\restoresymbol{TXF}{div}
\usepackage{tikz}
\usepackage{tikz-3dplot}
\usepackage{xcolor}
\usetikzlibrary{patterns, decorations.markings, arrows}
\usetikzlibrary{arrows.meta}
\usepackage{mathtools}
\usepackage[export]{adjustbox}
\usepackage{tcolorbox}
\usepackage{soul}

\usetikzlibrary{snakes}
\usetikzlibrary{calc}
\usetikzlibrary{decorations}

\tikzset{
  rarrow/.style={
    decoration={,
      markings,
      mark=at position 0.65 with {\arrow{latex}}},
    postaction={decorate}
  },
  larrow/.style={
    decoration={,
      markings,
      mark=at position 0.55 with {\arrow{latex reversed}}
    },
    postaction={decorate}
  }
}
\definecolor{mygreen}{RGB}{0,100,0}

\title{Kinematics, Cluster Algebras and Feynman Integrals}
\date{\today}
\author[a,b,c,d,e]{Song He,}
\author[a,d]{Zhenjie Li,} 
 \author[a,d]{Qinglin Yang}%
\affiliation[a]{CAS Key Laboratory of Theoretical Physics, Institute of Theoretical Physics, Chinese Academy of Sciences, Beijing 100190, China}
\affiliation[b]{
School of Fundamental Physics and Mathematical Sciences, Hangzhou Institute for Advanced Study, UCAS, Hangzhou 310024, China}
\affiliation[c]{ICTP-AP
International Centre for Theoretical Physics Asia-Pacific, Beijing/Hangzhou, China}
\affiliation[d]{School of Physical Sciences, University of Chinese Academy of Sciences, No.19A Yuquan Road, Beijing 100049, China}
\affiliation[e]{Peng Huanwu Center for Fundamental Theory, Hefei, Anhui 230026, P. R. China}
\emailAdd{songhe@itp.ac.cn}
\emailAdd{lizhenjie@itp.ac.cn}
\emailAdd{yangqinglin@itp.ac.cn}

\abstract{We identify cluster algebras for planar kinematics of conformal Feynman integrals in four dimensions, as sub-algebras of cluster algebras for Grassmannian $G(4,n)$ corresponding to $n$-point massless kinematics. We classify such algebras for cases through eight points, and provide evidence that singularities of the corresponding Feynman integrals are given by cluster variables and their algebraic generalizations. By sending a point to infinity, our results have implications for symbology of non-conformal Feynman integrals. We also find that dimensional reduction is achieved by {\it folding} and the resulting cluster algebras encode singularities of Feynman integrals in three dimensions. As a highly-nontrivial application, we study a eight-point three-loop wheel integral whose kinematics correspond to two-mass-easy box in the non-conformal limit: in addition to nine rational letters that are cluster variables, we also find three algebraic letters containing a new square root, in accordance with results from differential equations for two-loop case. Based on this alphabet, we bootstrap its symbol, which turns out to be strongly constrained by cluster adjacency conditions.
}

\begin{document}

\maketitle

\section{Introduction}

Recent years, tremendous progress has been made in unraveling mathematical structures of scattering amplitudes, especially in ${\cal N}{=}4$ supersymmetric Yang-Mills theory (sYM) in the planar limit. Due to the advanced symmetric structure, dual conformal invariance (DCI) and the Yangian invariance \cite{Drummond:2008vq,Drummond:2009fd}, its scattering amplitudes enjoy elegant mathematical and physical properties ({\it c.f.}~\cite{Arkani-Hamed:2016byb, Arkani-Hamed:2013jha}). Most recently, cluster algebras~\cite{fomin2002cluster,fomin2003cluster,fomin2007cluster} have played an important role in the studies of its scattering amplitudes, not only for its all-loop integrand~\cite{Arkani-Hamed:2016byb} but also for the functions after integration. The six- and seven-gluon amplitudes have been bootstrapped to impressively high loop orders~\cite{Dixon:2011pw,Dixon:2014xca,Dixon:2014iba,Drummond:2014ffa,Dixon:2015iva,Caron-Huot:2016owq,Dixon:2016nkn,Drummond:2018caf, Caron-Huot:2019vjl,Caron-Huot:2019bsq,Dixon:2020cnr,Golden:2021ggj,Caron-Huot:2020bkp}, and the starting point is the conjecture that their {\it symbol alphabet}~\cite{Goncharov:2010jf,Duhr:2011zq} is given by finite cluster algebras for Grassmannian $G(4,n)$ with $n=6,7$ respectively~\cite{Golden:2013xva,Golden:2014xqa}. Furthermore, great restriction has been imposed on the space of possible multiple polylogarithmic functions (MPL) due to the (extended) Steinmann relations~\cite{Steinmann1960a,Steinmann1960b, Caron-Huot:2016owq}
, which are closely related to the so-called {\it cluster adjacency conditions}~\cite{Drummond:2017ssj, Drummond:2018caf,Drummond:2018dfd}. Staring eight points, the cluster algebra becomes infinite and generally the symbol alphabet involves algebraic letters that go beyond cluster variables. Recent computations of two- and three-loop amplitudes~\cite{Zhang:2019vnm,He:2020vob,Li:2021bwg} based on the $\bar{Q}$-anomaly equation~\cite{CaronHuot:2011kk, CaronHuot:2010ek, Caron-Huot:2013vda} has provided new data for $n\geq 8$, which has led to novel mathematical structures directly related to cluster algebras and positivity~\cite{Arkani-Hamed:2019rds, Drummond:2019qjk,Henke:2019hve,Drummond:2019cxm,Henke:2021avn,Ren:2021ztg,Mago:2020kmp,He:2020uhb,Mago:2020nuv,Mago:2021luw}.

On the other hand, ${\cal N}{=}4$ sYM is also an extremely fruitful playground for new methods of evaluating Feynman integrals. There has been significant progress in computing and studying finite, conformal Feynman integrals contributing to amplitudes~\cite{Drummond:2006rz, Drummond:2007aua,Drummond:2010cz,ArkaniHamed:2010gh,Spradlin:2011wp,DelDuca:2011wh,Bourjaily:2013mma,Henn:2018cdp,Herrmann:2019upk, Bourjaily:2018aeq,Bourjaily:2019hmc,He:2020uxy,He:2020lcu}. Among all these studies, remarkably, tools of cluster algebras and adjacency properties also widely apply to individual Feynman integrals in ${\cal N}=4$ sYM for six and seven points~\cite{Caron-Huot:2018dsv, Drummond:2017ssj,Henn:2018cdp}, as well as more than eight points~\cite{He:2021esx, He:2021non, He:2021mme}).  These structures have been used to perform cluster bootstrap calculation for individual Feynman integrals as well \cite{Drummond:2017ssj,Henn:2018cdp,He:2021fwf,He:2021non}. Furthermore, cluster algebraic structures have also been found for more general Feynman integrals~\cite{Chicherin:2020umh} and for other quantities such as form factors~\cite{Dixon:2020bbt,Dixon:2021tdw}, indicating broader underlying connections between Feynman integrals and cluster algebras.

Especially motivated by~\cite{Chicherin:2020umh}, in this paper we will systematically explore the potential connection between conformal Feynman integrals in $\mathcal{N}{=}4$ sYM theory and cluster algebras. In particular, starting from the mathematical structure of the $G(4,n)$ cluster algebra for scattering amplitudes, we will assign a corresponding cluster algebra structure to conformal Feynman integrals by identifying their kinematics with appropriate $G(4,n)$ sub-algebras. The construction of such sub-algebras is based on finding sub-quivers of $G(4,n)$ that precisely control the kinematic variables associated with the Feynman integrals. We will explicitly provide all sub-algebras corresponding to kinematics with up to eight points and describe the general algorithm for obtaining them. 

Through the definition of these “kinematic quivers,” we will be able to determine the corresponding cluster alphabets for given kinematic configurations of specific Feynman integrals. For the six- and seven-point cases, these alphabets are believed to be sufficient to describe all symbol letters of the Feynman integrals. Starting from eight points, counterexamples begin to appear. Nevertheless, in the example discussed in this paper, although there exist symbol letters beyond the cluster variables, the associated singularity structure is still controlled by the cluster variables. This allows us to obtain the complete alphabet and to carry out the bootstrap for a particular three-loop integral.

Finally, we discuss two important implications of this construction. First, we find that restricting external legs to a three-dimensional subspace corresponds, at the level of cluster algebras, to a folding of the cluster algebra. Second, by fixing a conformal frame, these results can also be related to Feynman integrals and their symbol alphabets in more general theories, which follows from the discussions in \cite{Chicherin:2020umh}.

\subsection{Review and Notations}

We first review the planar kinematics of $n$ massless momenta, or equivalently an $n$-gon in (dual) space whose vertices are light-like separated, where the dual points are defined via 
\begin{equation}
p_i^\mu=x^\mu_i-x^\mu_{i{+}1},\ i=1,\cdots, n\ \  \text{(mod $n$)}
\end{equation} 
satisfying $x_{i,i{+}1}^2=0$. Such a kinematics can be described in terms of $n$ {\it momentum twistor} variables~\cite{Hodges:2009hk}, $Z^a_{i=1, \cdots, n}$ for $a=1,\cdots,4$, one for each massless momentum (null ray of the $n$-gon), defined by
\begin{equation}
	Z_i^I = \begin{pmatrix}
		\lambda_i^\alpha \\
		(x_i \lambda_i)^{\dot{\beta}}
	\end{pmatrix} \, , \qquad I\in\{1,\dots,4\} \, , \, {\alpha,\dot\beta} \in \{1,2\} \, .
\end{equation}
where the $\lambda_i^{\alpha}$ are the spinor-helicity variables defined via the map
\begin{equation}
p_i^\mu \rightarrow p_i^{\alpha \dot{\beta}} \equiv (p_i^\mu \sigma_{\mu})^{\alpha\dot{\beta}}=\lambda_i^{\alpha}\widetilde{\lambda}_i^{\dot{\beta}} \, ,
\label{eq:spinHel}
\end{equation}
by contracting $p_i^\mu$ with the four-vector of Pauli matrices $\sigma^\mu = (\mathbb{1}, \vec\sigma)$. These variables follow from expressing a dual point $x^\mu \in \mathbb{R}^{3,1}$ as a null projective vector $X^I \in \mathbb{R}^{4,2}$ satisfying $X^2=0$ and $X^I \sim T X^I$; after complexifying, such a 6d vector can be viewed as an antisymmetric tensor $X^{a,b}$ for SL$(4, \mathbb{C})$, whose fundamental representation is the twistor $Z^a$. Each point $x_i^\mu$ corresponds to a line given by the pair $(Z_{i{-}1}, Z_i)$ and since $Z_i \sim t_i Z_i$, they form homogeneous coordinates of $\mathbb{CP}^3$. %Conformal transformations in $x$ corresponds to SL(4) transformations of $Z$ (rotations of $X$), thus 
The space of DCI kinematics~\cite{Drummond:2006rz,Drummond:2007aua} is simply the Grassmannian mod torus action, $G(4,n)/T$ with dimension $3(n{-}5)$: conformal invariant quantities are torus-invariant functions, {\it i.e.} cross-ratios of Pl\"ucker coordinates 
\begin{equation}
\langle i j k l\rangle:={\rm det}(Z_i Z_j Z_k Z_l)
\end{equation} 
In particular, we have $x_{i,j}^2=\langle i{-}1 i j{-}1 j\rangle/(\langle i{-}1 i I_\infty\rangle \langle j{-}1 j I_\infty\rangle)$, where the factors involving the point (line) at infinity, $I_{\infty}$,
\begin{equation}
    I_{\infty}:=\left(\begin{matrix}
        0&0 & 1 &0 \\0&0 &0 & 1
    \end{matrix}\right) \,.
\end{equation}
It drops out in any DCI quantities. 

\begin{figure}[t]
    \centering
    \includegraphics[width=0.4\linewidth]{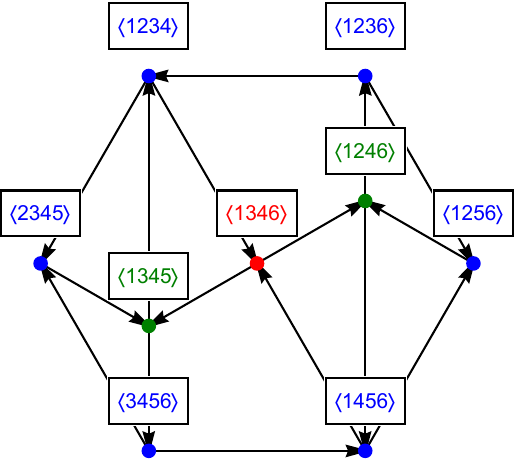}
    \caption{$G(4,6)$ and its sub-quiver}
    \label{fig:6pt}
\end{figure}

We are interested in {\it positive kinematics}, $G_{\geq 0} (4,n)/T$ where all (ordered) Pl\"uckers are non-negative, and it is natural to associate it with a {\it Grassmannian cluster algebra}~\cite{speyer2004tropical, speyer2005tropical}. For six, seven and eight points, the cluster algebras for $G(4,n)$ are of type $A_3, E_6$ and $E_7^{(1,1)}$ respectively~\footnote{The first two are finite types, and the last one is an infinite but finite mutation type; for $n>8$ they are infinite mutation type.}. For example, for six-pint case,  one can choose the initial quiver shown in Fig.~\ref{fig:6pt} for $G(4,6)$, where in addition to frozen variables in blue, we have $3(n{-}5)=3$ unfrozen (mutable) ${\cal A}$-coordinates forming an $A_3$ Dynkin diagram; by mutating them we have $9$ cluster variables for $A_3$; in this case the total $9+6$ ${\cal A}$ coordinates are exactly the $15$ Pl\"uckers $\langle i j k l\rangle$, which can form $9$ DCI combinations (up to multiplication) known as the {\it letters} of $A_3$ kinematics. Equivalently these $9$ DCI letters can be obtained as ${\cal X}$ coordinates, expressed in terms of $9$ {\it positive polynomials} of $f_1, f_2, f_3$ (${\cal X}$ coordinates of the initial quiver)
\begin{equation}
    f_1=\frac{\langle 1234\rangle\langle 3456\rangle}{\langle 2345\rangle\langle 1346\rangle},\ f_2=\frac{\langle 1345\rangle\langle 1246\rangle}{\langle 1234\rangle\langle 1456\rangle},\ f_3=\frac{\langle 1236\rangle\langle 1456\rangle}{\langle 1256\rangle\langle 1346\rangle}
\end{equation}
This provides a first example of parametrizing the kinematics and symbol letters in terms of cluster variables.

In this work, we will focus on general planar, conformal kinematics, {\it e.g.} for Feynman integrals with ``massive" corners (each consists of a pair of legs), which only depends on a subset of the $n$ dual points. One can arrive at such a kinematics by removing (non-adjacent) dual points from $x_1, x_2, \cdots, x_n$: {\it e.g.} putting $p_2$ and $p_3$ on a massive corner is equivalent to removing $x_3$, and a massive corner with $p_4, p_5$ amounts to removing $x_5$. The resulting kinematics are labelled by $(x_{i_1}, x_{i_2}, \cdots, x_{i_m})$ ({\it e.g.} $\{\cdots, x_2, x_4, x_6, \cdots\}$ in the above example), where $i_a$ and $i_{a{+}1}$ are separated by at most $2$, which correspond to $n$-point $m$-gon with $ n/2 \leq m\leq n$. We denote such a kinematics by $(n,m)$ with $A,B,\cdots$ distinguishing configurations of massive corners. Generically the dimension of a $n$-point $m$-gon kinematics is given by $d=n{+}2m{-}15$, except for the special, four-mass box case where $d=2$. A priori it is unclear at all if one can find any sub-algebra of the $G(4,n)$ cluster algebra which parametrizes such a kinematics. In this work, we show that it is indeed the case by going through all kinematics with $n\leq 8$, which we expect to work for higher $n$ as well.

\section{Cluster algebras for conformal Feynman integrals}

\subsection{Kinematic quivers and DCI kinematics with massive corners}

To find a suitable sub-quiver of $G(4,n)$ to represent a $(n,m)$ kinematics,  the main idea is to {\it freeze} some mutable variables, such that the remaining sub-quiver with $d$ mutable nodes becomes independent of the removed dual points. For kinematics with one massive corner, say, $(23)$, we should identify a codimension two sub-algebra, which is independent of dual point $x_{3}$. We would like to therefore ``delete" the frozen variables $\langle 123 n \rangle\propto x^2_{1,3}$ and $\langle 2345\rangle\propto x^2_{3,5}$ that depend on $x_3$, which can be achieved if they only connect to other frozen variables; %since we can freely add or delete arrows between frozen variables without  affecting its cluster algebra. 
this motivates us to find a quiver where each of them is only connected to a unique {\it mutable} node, and the desired sub-quiver is obtained by freezing these two nodes.

\begin{figure}[t]
    \centering
      \begin{tikzpicture}[scale=0.2]
        \draw[black,thick] (0,0)--(5,0)--(6.55,4.76)--(2.50,7.69)--(-1.55,4.76)--cycle;
        \draw[black,thick] (1.5,9.43)--(2.5,7.69)--(3.5,9.43);
        \draw[black,thick] (-1.21,-1.99)--(0,0);
        \draw[black,thick] (5,0)--(6.21,-1.99);
         \draw[black,thick] (6.55,4.76)--(8.45,6.37);
        \draw[black,thick] (-3.55,6.37)--(-1.55,4.76);
        \filldraw[black] (1.5,9.43) node[anchor=south] {{$2$}};
\filldraw[black] (3.5,9.43) node[anchor=south] {{$3$}};
\filldraw[black] (8.45,6.37) node[anchor=west] {{$4$}};
\filldraw[black] (6.21,-1.99) node[anchor=north] {{$5$}};
\filldraw[black] (-1.21,-1.99) node[anchor=north] {{$6$}};
\filldraw[black] (-3.55,6.37) node[anchor=east] {{$1$}};
    \end{tikzpicture}\quad
    \begin{tikzpicture}[scale=0.2]
        \draw[black,thick] (0,0)--(0,4)--(3.46,6)--(6.93,4)--(6.93,0)--(3.46,-2)--cycle;
        \draw[black,thick] (6.93,4)--(8.66,5);
        \draw[black,thick] (0,4)--(-1.73,5);
        \draw[black,thick] (3.46,-2)--(3.46,-4);
        \filldraw[black] (-1.73,5) node[anchor=south east] {{$1$}};
        \filldraw[black] (2.26,7.73) node[anchor=south] {{$2$}};
        \filldraw[black] (4.66,7.73) node[anchor=south] {{$3$}};
        \filldraw[black] (8.66,5) node[anchor=south west] {{$4$}};
        \filldraw[black] (8.13,-1.53) node[anchor=north west] {{$5$}};
        \filldraw[black] (3.46,-4) node[anchor=north] {{$6$}};
        \filldraw[black] (-1.2,-1.53) node[anchor=north east] {{$7$}};
        \draw[black,thick] (2.46,7.73)--(3.46,6)--(4.46,7.73);
        \draw[black,thick] (0,0)--(-1.2,-1.53);
        \draw[black,thick] (6.93,0)--(8.13,-1.53);
        %\filldraw[black] (3.46,2) node[anchor=center] {{6D}};
    \end{tikzpicture}\quad
    \begin{tikzpicture}[scale=0.2]
        \draw[black,thick] (0,0)--(5,0)--(6.55,4.76)--(2.50,7.69)--(-1.55,4.76)--cycle;
        \draw[black,thick] (1.5,9.43)--(2.5,7.69)--(3.5,9.43);
        \draw[black,thick] (-1.21,-1.99)--(0,0);
        \draw[black,thick] (6.83,-0.81)--(5,0)--(5.21,-1.99);
         \draw[black,thick] (6.55,4.76)--(8.45,6.37);
        \draw[black,thick] (-3.55,6.37)--(-1.55,4.76);
        \filldraw[black] (1.5,9.43) node[anchor=south] {{$2$}};
\filldraw[black] (3.5,9.43) node[anchor=south] {{$3$}};
\filldraw[black] (8.45,6.37) node[anchor=west] {{$4$}};
\filldraw[black] (6.83,-0.81) node[anchor=west] {{$5$}};
\filldraw[black] (5.21,-1.99) node[anchor=north] {{$6$}};
\filldraw[black] (-1.21,-1.99) node[anchor=north] {{$7$}};
\filldraw[black] (-3.55,6.37) node[anchor=east] {{$1$}};
    \end{tikzpicture}\quad
    \begin{tikzpicture}[scale=0.2]
        \draw[black,thick] (0,0)--(5,0)--(6.55,4.76)--(2.50,7.69)--(-1.55,4.76)--cycle;
        \draw[black,thick] (2.5,7.69)--(2.5,9.43);
        \draw[black,thick] (-0.21,-1.99)--(0,0)--(-1.83,-0.81);
        \draw[black,thick] (6.83,-0.81)--(5,0)--(5.21,-1.99);
         \draw[black,thick] (6.55,4.76)--(8.45,6.37);
        \draw[black,thick] (-3.55,6.37)--(-1.55,4.76);
        \filldraw[black] (2.5,9.43) node[anchor=south] {{$2$}};
\filldraw[black] (8.45,6.37) node[anchor=west] {{$3$}};
\filldraw[black] (6.83,-0.81) node[anchor=west] {{$4$}};
\filldraw[black] (5.21,-1.99) node[anchor=north] {{$5$}};
\filldraw[black] (-0.21,-1.99) node[anchor=north] {{$6$}};
\filldraw[black] (-1.83,-0.81) node[anchor=east] {{$7$}};
\filldraw[black] (-3.55,6.37) node[anchor=east] {{$1$}};
    \end{tikzpicture}
    \caption{kinematics for six and seven points; for the two pentagon kinematics at seven-point, the first is $(7,5)$ A and the second $(7,5)$ B}
    \label{fig:67ptkinemaics}
\end{figure}

Let us begin with the simplest examples as illustration. In Fig. \ref{fig:67ptkinemaics} we list all six- and seven-point kinematics with massive corners. According to our convention, they are called (6,5) for the six-point kinematics, (7,6) for the first seven-point kinematics, and (7,5) A, (7,5) B for the two two-mass seven-point kinematics. Firstly, for the $(6,5)$ kinematics, we choose the quiver of $G(4,6)$ as in Fig.~\ref{fig:6pt}, and freeze $\langle 1345\rangle$ and $\langle 1246\rangle$, which are connected to $\langle 1236\rangle$ and $\langle 2345\rangle$ respectively. The resulting quiver has only one mutable variable $\langle 1346\rangle$, and four frozen ones connected to it, which forms an $A_1$ sub-algebra of $A_3$ with no $x_3$ dependence! In terms of all possible two $\mathcal X$-coordinates, we have 
\begin{equation}
f_2=\frac{\langle 1345\rangle\langle 1246\rangle}{\langle 1234\rangle\langle 1456\rangle}, 1+f_2=\frac{\langle 1245\rangle \langle 1346\rangle}{\langle 1234\rangle\langle 1456\rangle}
\end{equation}
both are independent of $x_3$. This quiver therefore describes the (6,5) kinematics.  %Similarly in Fig.~\ref{fig:7pt} we identify a $D_4$ sub-algebra by freezing $\langle 1345\rangle$ and $\langle 1247 \rangle$, for $(7,6)$ kinematics in Fig.~\ref{fig:67ptkinemaics}. 

\begin{figure}[t]
    \centering
    \includegraphics[width=0.4\linewidth]{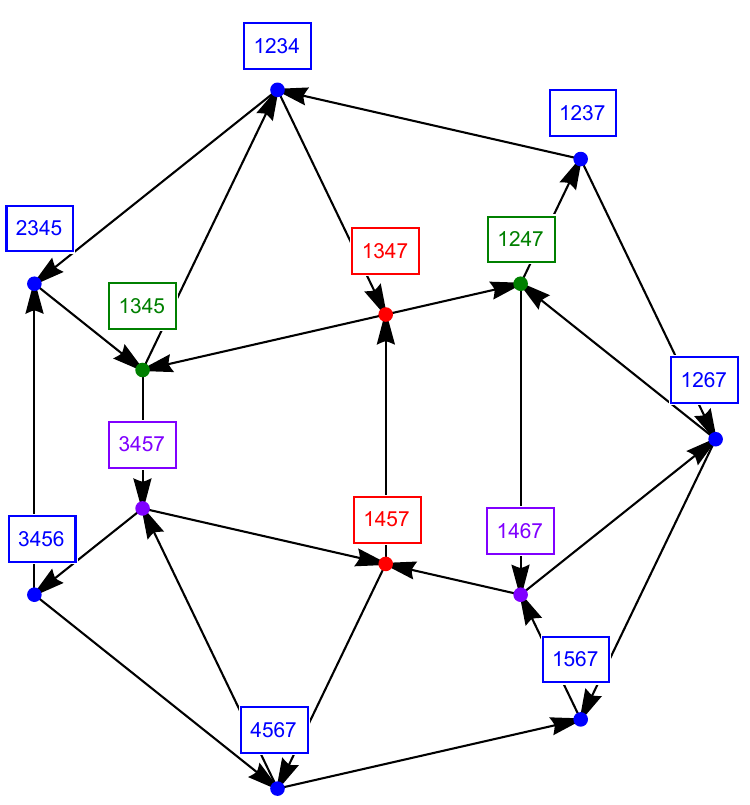}
    \caption{$G(4,7)$ and its different sub-quivers}
    \label{fig:7pt}
\end{figure}

Moving to seven-point kinematics, we select the quiver as Fig.\ref{fig:7pt}. To achieve (7,6) kinematics, we identify a $D_4$ sub-algebra by freezing $\langle 1345\rangle$ and $\langle 1247 \rangle$ (the green ones in the figure), since it is independent of $\langle7123\rangle$,$\langle2345\rangle$ by deleting the $x_3$ dual point. On top of that, kinematics $(7,5)$ A has no dependence on dual point $x_6$. Therefore, we should freeze two more nodes $\langle3457\rangle$ and $\langle1467\rangle$ to get its kinematic quiver, which is an $A_2$ sub-algebra. In Fig. \ref{fig:7pt} we show how to freeze two (four) nodes of $E_6$ to arrive at these sub-algebras.

%as we have discussed $(7,6)$ kinematics corresponds to a $D_4$ sub-algebra and $(7,5)$ A corresponds to an $A_2$: here the hexagon is independent of $\langle7123\rangle$,$\langle2345\rangle$, and the pentagon further independent of $\langle3456\rangle$ and $\langle5671\rangle$. In Fig. \ref{fig:7pt} we show how to freeze two (four) nodes of $E_6$ to arrive at these sub-algebras.  

In general, if none of the $n{-}m$ massive corners are adjacent to each other, it is always possible to find a quiver containing all these $2(n{-}m)$ mutable variables and we reach at our sub-quiver by freezing all of them! %For example, to go to $(7,5)$ A kinematics (see Fig.~\ref{fig:67ptkinemaics} and Fig.~\ref{fig:7pt}), we just freeze two more from that of $D_4$ (for $(7,6)$), $\langle 3457 \rangle$ and $\langle 1567 \rangle$, which gives a $A_2$ sub-algebra. 
For such cases, this algorithm of freezing nodes is equivalent to that of \cite{He:2021non}, where we go to lower-dimensional positroid cells by setting unwanted frozen variables to zero: the sub-quiver we obtain is simply the dual graph of the plabic graph for the cell, whose face variables correspond to $\mathcal X$-coordinates of the sub-algebra! Such examples include $D_d$ sub-algebras of  $G(4,d{+}3)/T$ for $d=4,5,6$, and the $D_{4,1}$ or affine $D_4$ of eight points ($(8,6)$ A kinematics), which are various hexagon kinematics with non-adjacent massive corners~\cite{He:2021non}. 

\begin{figure}[t]
    \centering
    \includegraphics[scale=0.7]{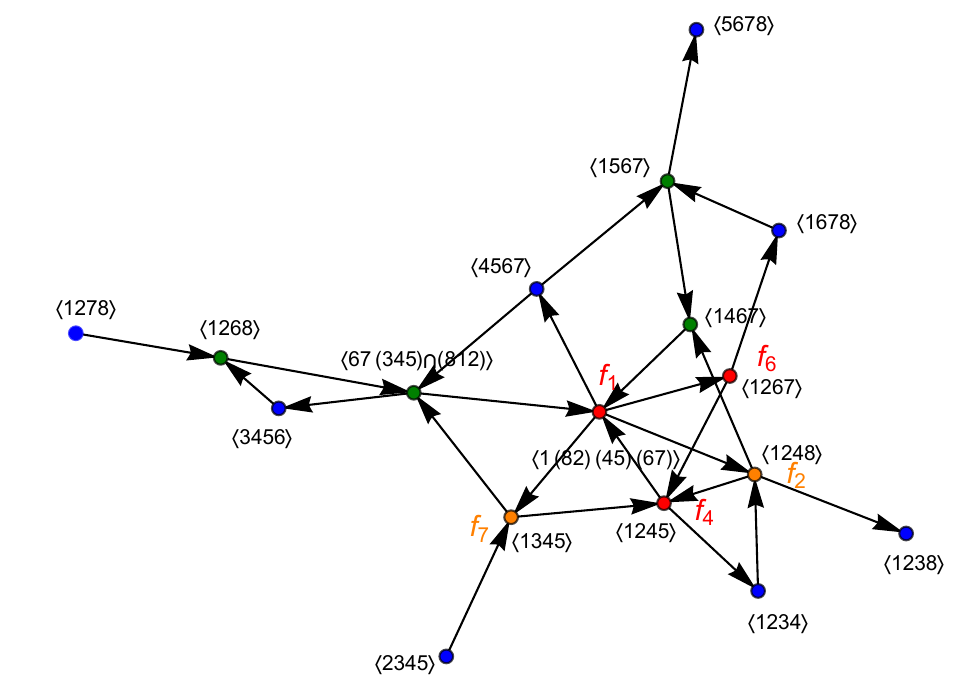}
    \caption{An initial quiver of $G(4,8)$, where the $D_{4,1}$ sub-quiver with red and orange nodes corresponds to the kinematics (8,6) C, and the $D_3$ sub-quiver with red ones corresponds to the kinematics (8,5) A, blue nodes are original frozen variables of $G(4,8)$. The rest unfrozen nodes of $G(4,8)$ are green.}
    \label{fig:D3quiver}
\end{figure}

However, for kinematics with adjacent massive corners like (7,5) B, we cannot set enough frozen variables to zero to reduce to the correct dimension, and general sub-quiver does not correspond to a positroid cell. There will be more such examples if we going to higher points, and in Table\ref{table} we summarize all eight-point kinematics. Take $(8,5)$ A kinematics as an example, where we need to remove $x_3, x_6, x_7$  (see Fig.~\ref{fig:pentagon} and Fig.\ref{fig:D3quiver}). As we have discussed in \cite{He:2021non}, we cannot parametrize this kinematics by a cell. Nevertheless, we can still find a quiver of $G(4,8)$ by freezing six nodes: in addition to $\langle 1345\rangle$ and $\langle 1248 \rangle$ for corner $(23)$ (which can be reached via a cell), we freeze $4$ more variables as shown in Fig.~\ref{fig:D3quiver}. The result is a $D_3$ sub-algebra \footnote{Although it is equivalent to, and usually written as $A_3$, to distinguish it from the six-point $A_3$ algebra, we always call it $D_3$ in this work}, where the three kinematic cross-ratios 
\begin{equation}
u=\frac{x_{1,7}^2 x_{2,5}^2}{x_{1,5}^2 x_{2,7}^2},\  v=\frac{x_{1,4}^2 x_{5,7}^2}{x_{1,5}^2 x_{4,7}^2},\  w=\frac{x_{1,5}^2 x_{2,4}^2}{x_{1,4}^2 x_{2,5}^2},
\end{equation} 
are parametrized in terms of the three $\mathcal X$-coordinates of the initial quiver:
\begin{equation}\label{1}
u=\frac1{1{+}f_6}, v=\frac{1{+}f_6{+}f_4f_6}{(1{+}f_4{+}f_1f_4)(1{+}f_6)}, w=\frac{1{+}f_6}{1{+}f_6{+}f_4f_6}\,.
\end{equation}
Note that since $(7,5)$ B in Fig.~\ref{fig:67ptkinemaics} is a degeneration of this kinematics with $\langle1234\rangle\to0$, we can freeze $\langle1245\rangle$ in this quiver to get a $A_2$, which is a different $A_2$ from that of $(7,5)$ A.

Finally, as shown in Fig.~\ref{fig:D3quiver}, this $D_3$ for $(8,5)$A is a sub-algebra of that for $(8,6)$C: if we do not freeze $\langle 1345\rangle$ and $\langle 1248 \rangle$, we find a $5$-dim sub-algebra, which is a $D_{4,1}$ (different from that of $(8,6)$A); it depends on $u, v, w$ as well as $u', v'$
\begin{align}
  u^\prime:=\frac{x_{1,3}^2x_{5,7}^2}{x_{1,5}^2x_{3,7}^2}{=}\frac{f_6}{(1{+}f_6)Y},v^\prime:=\frac{x_{1,7}^2x_{3,5}^2}{x_{1,5}^2x_{3,7}^2}{=}\frac{f_1f_2f_4f_7}{(1{+}f_6)Y}
\end{align}
where in addition to $f_1, f_4, f_6$ we have two more initial ${\cal X}$ coordinates $f_2$ and $f_7$, and we have defined $Y:=1{+}f_1{+}f_1f_2{+}f_1f_7{+}f_1f_2f_7{+}f_1f_2f_4f_7$.

\begin{table}[htbp]
\centering
\begin{tabular}{|c|c|c|c|c|}
\hline
(n,m)   & kinematics   & initial quiver & CA & folding            \\ \hline
(8,8)   &   \begin{tikzpicture}[scale=0.11]
        \draw[black,thick] (0,0)--(4,0)--(6,2.46)--(6,5.46)--(4,7.93)--(0,7.93)--(-2,5.46)--(-2,2.46)--cycle;
       \draw[black,thick] (4,7.93)--(5,9.66);
        \draw[black,thick] (4,0)--(5,-1.73);
        \filldraw[black] (5,-1.73) node[anchor=north west] {{$4$}};
        \filldraw[black] (7.73, 2.26) node[anchor=north west] {{$3$}};
        \filldraw[black] (7.73, 4.66) node[anchor=south west] {{$2$}};
        \filldraw[black] (5,8.66) node[anchor=south west] {{$1$}};
        \filldraw[black] (-1,8.66) node[anchor=south east] {{$8$}};
        \filldraw[black] (-3.73,2.46) node[anchor=north east] {{$6$}};
        \filldraw[black] (-3.73,4.46) node[anchor=south east] {{$7$}};
        \filldraw[black] (-1,-1.73) node[anchor=north east] {{$5$}};
        \draw[black,thick] (7.73,1.46)--(6,2.46);
        \draw[black,thick] (6,5.46)--(7.73,6.46);
        \draw[black,thick] (-3.73,1.46)--(-2,2.46);
         \draw[black,thick] (-2,5.46)--(-3.73,6.46);
        \draw[black,thick] (0,0)--(-1,-1.73);
        \draw[black,thick] (0,7.93)--(-1,9.66);
        %\filldraw[black] (3.46,2) node[anchor=center] {{6D}};
    \end{tikzpicture}                 &  \includegraphics[scale=1.0]{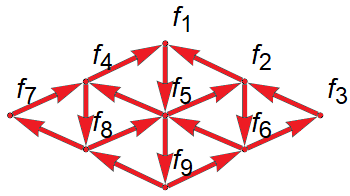}              & $E_{7}^{(1,1)}$ & name unknown\\ \hline
(8,7)   &        \begin{tikzpicture}[scale=0.14]
        \draw[black,thick]  (0,4)-- (3.127,2.494)-- (3.900,-0.890)-- (1.736,-3.604)-- (-1.736,-3.604)--(-3.900,-0.890)-- (-3.127,2.494)--cycle;
    \draw[black,thick](-1,5)--(0,4)--(1,5);
    \draw[black,thick](3.127,2.494)--(4.127,3);
    \draw[black,thick](-3.127,2.494)--(-4.127,3);
     \draw[black,thick](3.900,-0.890)--(4.900,-1);
      \draw[black,thick](-3.900,-0.890)--(-4.900,-1);
       \draw[black,thick] (1.736,-3.604)-- (2.736,-4.204);
        \draw[black,thick](-1.736,-3.604)-- (-2.736,-4.204);
        \filldraw[black] (-4.127,3) node[anchor=south east] {{$1$}};
        \filldraw[black](-1,5) node[anchor=south] {{$2$}};
        \filldraw[black] (1,5) node[anchor=south] {{$3$}};
        \filldraw[black](4.127,3)  node[anchor=south west] {{$4$}};
        \filldraw[black](4.900,-1) node[anchor=west] {{$5$}};
        \filldraw[black](2.736,-4.204) node[anchor=north west] {{$6$}};
        \filldraw[black] (-2.736,-4.204) node[anchor=north east] {{$7$}};
        \filldraw[black] (-4.900,-1) node[anchor=east] {{$8$}};
    \end{tikzpicture}               &     \includegraphics[scale=0.8]{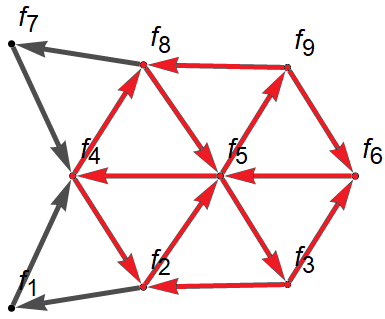}          & $E_{6,1}$  & $F_{4,1}$ \\ \hline
(8,6)A & \begin{tikzpicture}[scale=0.13]
        \draw[black,thick] (0,0)--(4,0)--(6,3.46)--(4,6.93)--(0,6.93)--(-2,3.4)--cycle;
       \draw[black,thick] (4,6.93)--(5,8.66);
        \draw[black,thick] (4,0)--(5,-1.73);
        \filldraw[black] (5,-1.73) node[anchor=north west] {{$4$}};
        \filldraw[black] (7.73, 2.26) node[anchor=north west] {{$3$}};
        \filldraw[black] (7.73, 4.66) node[anchor=south west] {{$2$}};
        \filldraw[black] (5,8.66) node[anchor=south west] {{$1$}};
        \filldraw[black] (-1,8.66) node[anchor=south east] {{$8$}};
        \filldraw[black] (-3.73,2.46) node[anchor=north east] {{$6$}};
        \filldraw[black] (-3.73,4.46) node[anchor=south east] {{$7$}};
        \filldraw[black] (-1,-1.73) node[anchor=north east] {{$5$}};
        \draw[black,thick] (7.73,2.46)--(6,3.46)--(7.73,4.46);
        \draw[black,thick] (-3.73,2.46)--(-2,3.46)--(-3.73,4.46);
        \draw[black,thick] (0,0)--(-1,-1.73);
        \draw[black,thick] (0,6.93)--(-1,8.66);
        %\filldraw[black] (3.46,2) node[anchor=center] {{6D}};
    \end{tikzpicture}            &      \includegraphics[scale=0.6]{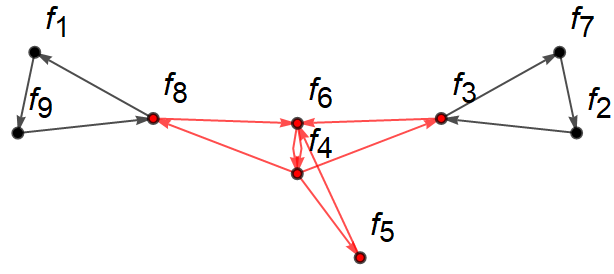}            & $D_{4,1}$      & $B_{3,1}$ \\ \hline
(8,6)B & \begin{tikzpicture}[scale=0.11]
        \draw[black,thick] (0,0)--(0,4)--(3.46,6)--(6.93,4)--(6.93,0)--(3.46,-2)--cycle;
        \draw[black,thick] (6.93,4)--(8.66,5);
        \draw[black,thick] (0,4)--(-1.73,5);
        \draw[black,thick] (3.46,-2)--(3.46,-4);
        \filldraw[black] (-1.73,5) node[anchor=south east] {{$1$}};
        \filldraw[black] (2.26,7.73) node[anchor=south] {{$2$}};
        \filldraw[black] (4.66,7.73) node[anchor=south] {{$3$}};
        \filldraw[black] (8.66,5) node[anchor=south west] {{$4$}};
        \filldraw[black] (8.93,0) node[anchor=west] {{$5$}};
        \filldraw[black] (7.93,-1.73) node[anchor=north] {{$6$}};
        \filldraw[black] (3.46,-4) node[anchor=north] {{$7$}};
        \filldraw[black] (-1.2,-1.53) node[anchor=north east] {{$8$}};
        \draw[black,thick] (2.46,7.73)--(3.46,6)--(4.46,7.73);
        \draw[black,thick] (0,0)--(-1.2,-1.53);
        \draw[black,thick] (8.93,0)--(6.93,0)--(7.93,-1.73);
        %\filldraw[black] (3.46,2) node[anchor=center] {{6D}};
    \end{tikzpicture}            &  \includegraphics[scale=0.8]{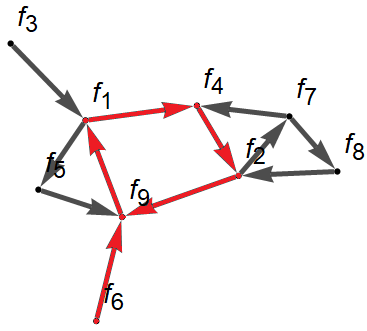}              & $D_5$       & $B_4$    \\ \hline
(8,6)C &  \begin{tikzpicture}[scale=0.11]
                \draw[black,thick] (0,0)--(4,0)--(6,3.46)--(4,6.93)--(0,6.93)--(-2,3.46)--cycle;
                \draw[black,thick] (0,6.93)--(-1,8.66);
                \draw[black,thick] (6,7.22)--(4,6.93)--(4.5,8.66);
                \draw[black,thick] (7.74,2.46)--(6,3.46)--(7.74,4.46);
                \draw[black,thick] (4,0)--(5,-1.73);
                \draw[black,thick] (0,0)--(-1,-1.73);
                \draw[black,thick] (-2,3.46)--(-4,3.46);
                \filldraw[black] (-1,8.66) node[anchor=south east] {{4}};
                \filldraw[black] (4.5,8.66) node[anchor=south west] {{5}};
                \filldraw[black] (6,7.22) node[anchor=south west] {{6}};
                \filldraw[black] (7.74,4.46) node[anchor=west] {{7}};
                \filldraw[black] (7.74,2.46) node[anchor=west] {{8}};
                \filldraw[black] (5,-1.73) node[anchor=north west] {{1}};
                \filldraw[black] (0,-1.73) node[anchor=north east] {{2}};
                \filldraw[black] (-4,3.46) node[anchor=east] {{3}};
            \end{tikzpicture}           &  \includegraphics[scale=0.5]{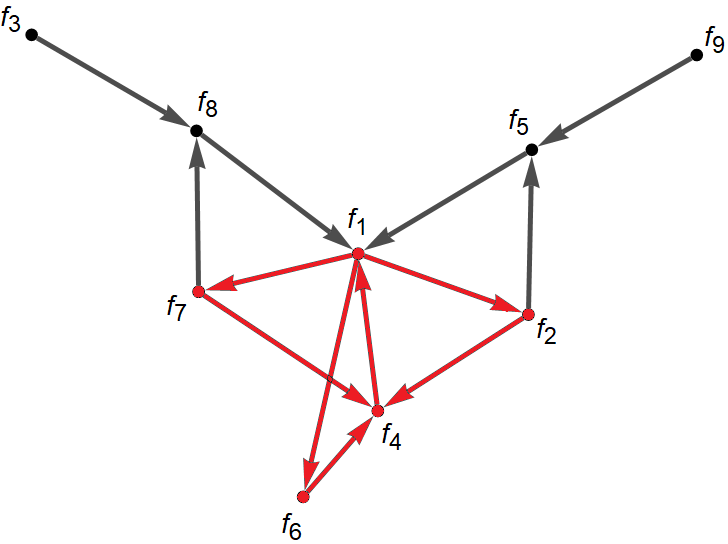}              & $D_{4,1}$  & $B_{3,1}$     \\ \hline
(8,5)A & \begin{tikzpicture}[scale=0.11]
        \draw[black,thick] (0,0)--(5,0)--(6.55,4.76)--(2.50,7.69)--(-1.55,4.76)--cycle;
        \draw[black,thick] (1.5,9.43)--(2.5,7.69)--(3.5,9.43);
        \draw[black,thick] (-0.21,-1.99)--(0,0)--(-1.83,-0.81);
        \draw[black,thick] (6.83,-0.81)--(5,0)--(5.21,-1.99);
         \draw[black,thick] (6.55,4.76)--(8.45,6.37);
        \draw[black,thick] (-3.55,6.37)--(-1.55,4.76);
        \filldraw[black] (1.5,9.43) node[anchor=south] {{$2$}};
\filldraw[black] (3.5,9.43) node[anchor=south] {{$3$}};
\filldraw[black] (8.45,6.37) node[anchor=west] {{$4$}};
\filldraw[black] (6.83,-0.81) node[anchor=west] {{$5$}};
\filldraw[black] (5.21,-1.99) node[anchor=north] {{$6$}};
\filldraw[black] (-0.21,-1.99) node[anchor=north] {{$7$}};
\filldraw[black] (-1.83,-0.81) node[anchor=east] {{$8$}};
\filldraw[black] (-3.55,6.37) node[anchor=east] {{$1$}};
    \end{tikzpicture}               &  \includegraphics[scale=0.5]{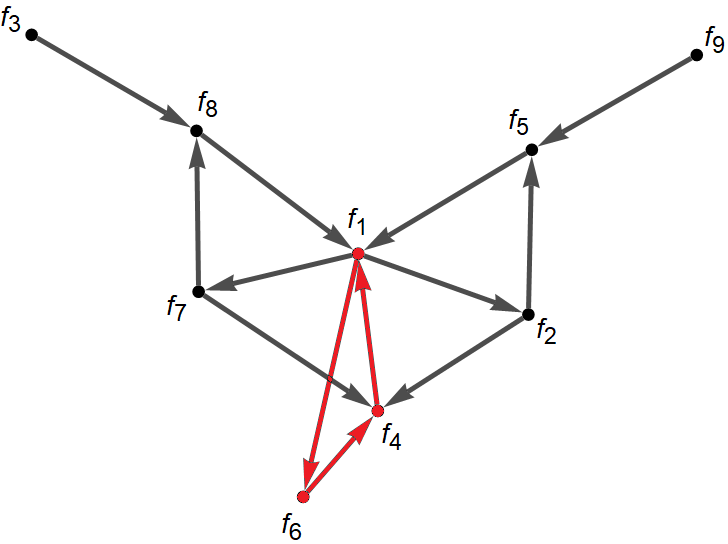}              & $D_3$  &          \\ \hline
(8,5)B &  \begin{tikzpicture}[scale=0.13]
        \draw[black,thick] (0,0)--(5,0)--(6.55,4.76)--(2.50,7.69)--(-1.55,4.76)--cycle;
        \draw[black,thick] (2.5,7.69)--(2.5,9.43);
        \draw[black,thick] (-0.21,-1.99)--(0,0)--(-1.83,-0.81);
        \draw[black,thick] (6.83,-0.81)--(5,0)--(5.21,-1.99);
         \draw[black,thick] (6.55,4.76)--(8.45,6.37);
        \draw[black,thick] (-3.45,6.37)--(-1.55,4.76)--(-3.65,4.37);
        \filldraw[black] (2.5,9.43) node[anchor=south] {{$8$}};
\filldraw[black] (8.45,6.37) node[anchor=west] {{$1$}};
\filldraw[black] (6.83,-0.81) node[anchor=west] {{$2$}};
\filldraw[black] (5.21,-1.99) node[anchor=north] {{$3$}};
\filldraw[black] (-0.21,-1.99) node[anchor=north] {{$4$}};
\filldraw[black] (-1.83,-0.81) node[anchor=east] {{$5$}};
\filldraw[black] (-3.65,4.37) node[anchor=east] {{$6$}};
\filldraw[black] (-3.45,6.37) node[anchor=south] {{$7$}};
    \end{tikzpicture}                &     \includegraphics[scale=0.6]{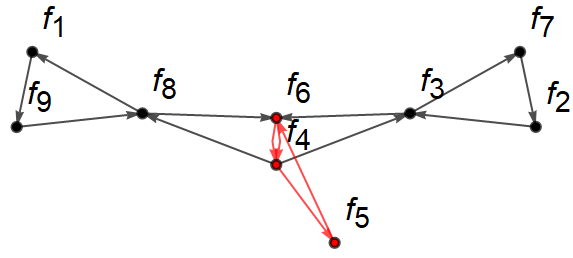}           & $A_{2,1}$  &     \\ \hline
\end{tabular}
\caption{Quivers and cluster algebras for eight-point kinematics}
\label{table}
\end{table}

In practice, for our algorithm of freezing nodes, it is more efficient to directly work with $\mathcal{X}$-coordinates as opposed to $\mathcal{A}$-coordinates. 
After some works in finding suitable initial quivers, it is straightforward to find sub-algebras for eight-point cases as we present in Table~\ref{table}. We locate their initial quivers as the red parts in the $G(4,8)$ quiver on the third column. Note that we choose the same initial quiver for $(8,6)$ A and $(8,5)$ B, and the same one for $(8,6)$ C and $(8,5)$ A, to emphasis the inclusion relations $A_{2,1}\subset D_{4,1}$(A) and $D_3\subset D_{4,1}$(C). For each sub-quiver of eight-point kinematics in the table, we record the expressions of their $\mathcal{X}$-coordinates $f_i$ in terms of Pl\"ucker coordinates in the ancillary file {\bf CA.m}.

%In general, we have to use the algorithm of freezing nodes of certain $G(4,n)$ quiver, and indeed we find sub-quivers for all kinematics of $n=8$ (see Table~\ref{table}). 

\subsection{Cluster variables from kinematic quivers and singularities}

Now we study implications of these cluster algebras for conformal Feynman integrals with the kinematics.

For six- and seven-point cases, all conformal integrals computed so far~\cite{ArkaniHamed:2010gh,Spradlin:2011wp,DelDuca:2011wh,Bourjaily:2018aeq,He:2020uxy,He:2020lcu,Henn:2018cdp,Herrmann:2019upk} are MPLs with symbol letters contained in $A_3$ and $E_6$ respectively. Even in these simpler cases, it is still non-trivial that the letters of multi-loop integrals with a given kinematics is captured by the corresponding sub-algebra, such as various seven-point ladder integrals with $(7,6)$ and $(7,5)$ A,B kinematics.

Let us look into the corresponding alphabets from kinematic quivers. We have $16+7$ $\mathcal{A}$-coordinates for the $D_4$ algebra in Fig.\ref{fig:7pt}: the unfrozen variables are
\begin{align}
&\bigl\{\langle 1(27)(34)(56)\rangle, 
 \langle 4(12)(35)(67)\rangle, 
 \langle 5(12)(34)(67)\rangle, 
 \langle 6(12)(34)(57)\rangle, 
 \langle 7(12)(34)(56)\rangle, 
 \nonumber \\
 &\langle 1245\rangle, \langle 1246\rangle,\langle 1256\rangle, \langle 1257\rangle, \langle 1346\rangle, \langle 1347\rangle, 
 \langle 1456\rangle, \langle 1457\rangle, \langle 1467\rangle, \langle 3457\rangle, 
 \langle 3467\rangle\bigr\}\,,
\end{align}
and the frozen ones are
\begin{equation}
\bigl\{\langle 1234\rangle, \langle 1247\rangle, \langle 1267\rangle, \langle 1345\rangle, 
 \langle 1567\rangle, \langle 3456\rangle, \langle 4567\rangle\bigr\}\,.
\end{equation}
%Besides the initial $7$ frozen + $4$ unfrozen $\mathcal{A}$-coordinates for $D_4$ in the quiver, we have $12$ more $\mathcal{A}$-coordinates
%\begin{align}
%\bigl\{\langle 1(27)(34)(56)\rangle, 
% \langle 4(12)(35)(67)\rangle, 
% \langle 5(12)(34)(67)\rangle,\langle 6(12)(34)(57)\rangle, 
% &\langle 7(12)(34)(56)\rangle, 
% \langle 1245\rangle, \langle 1246\rangle,\\ 
% &\langle 1256\rangle, \langle 1257\rangle, \langle 1346\rangle,  
% \langle 1456\rangle,  \langle 3467\rangle\bigr\}
%\end{align}
Among these, alphabet of $(7,5)$ A ($A_2$ algebra) contains 
\begin{equation}
\bigl\{\langle 4(12)(35)(67)\rangle,\langle 1245\rangle,
 \langle 1347\rangle, \langle 1457\rangle,  \langle 3467\rangle\bigr\}
 \end{equation}
as unfrozen variables and six frozen variables which are
\begin{equation}
\bigl\{\langle 1234\rangle, \langle 1247\rangle, \langle 1345\rangle, \langle1467\rangle,\ \langle3457\rangle,\ 
\langle 4567\rangle\bigr\}\,.
\end{equation}
These alphabets can be summarized as the tower $E_6 \supset D_4 \supset ({\rm two})~A_2$.

Finally, the $(7,5)$ B kinematics, as we mentioned, is the limit $\langle1234\rangle\to0$ (or $w\to0$) of $(8,5)$ A kinematics. For $(8,5)$A case, the $D_3$ sub-algebra has nine unfrozen $\mathcal A$-coordinates
\begin{align}
\bigl\{\langle 1 (28)(34)(67)\rangle,\, 
\langle 1 (28)(45)(67)\rangle,\, 
&\langle 4 (18)(35)(67)\rangle,\, 
\langle 4 (12)(35)(67)\rangle,\nonumber\\
&\langle 1267\rangle,
\langle 3467\rangle,
\langle 1348\rangle,
\langle 1458\rangle,
\langle 1245\rangle\bigr\}.
\end{align}
and seven frozen ones
\begin{equation}
\bigl\{\langle 812(67)\cap(345)\rangle,\langle 6781\rangle,\langle 1467\rangle,\langle 4567\rangle,\langle 8124\rangle,\langle 1234\rangle,\langle 1345\rangle\bigr\}\,.
\end{equation} 
Alphabet of (7,5) B contains $5+6$ $\mathcal{A}$-coordinates from them: five unfrozen variables
\begin{equation}\{\langle1(27)(34)(56)\rangle,\langle3(17)(24)(56)\rangle,\langle1256\rangle,\langle1347\rangle,\langle2356\rangle\}\end{equation} 
and six frozen ones
\begin{equation}\{\langle2(17)(34)(56)\rangle,\langle1234\rangle,\langle1567\rangle,\langle1237\rangle,\langle1356\rangle,\langle3456\rangle\}\,,\end{equation}
which form $5$ DCI combinations $\{u,v,1-u,1-v,1-u-v\}$ with 
\begin{equation}u=\frac{x_{2,4}^2x_{6,1}^2}{x_{2,6}^2x_{1,4}^2},\ v=\frac{x_{1,3}^2x_{4,6}^2}{x_{1,4}^2x_{3,6}^2}.
\end{equation} 
All these results meet exactly the alphabets of Feynman integrals in, {\it e.g.}, \cite{Caron-Huot:2018dsv,Bourjaily:2018aeq,He:2020uxy,He:2020lcu,He:2021fwf,Henn:2018cdp}.
%One sees that at $x_6\to\infty$, $\frac{x_{i,6}^2}{x_{j,6}^2}\to1$ for arbitrary $i$ and $j$. Therefore $u\to \frac{x_{2,4}^2}{x_{1,4}^2}:=z_2$, $v\to \frac{x_{1,3}^2}{x_{1,4}^2}:=z_1$ and we finally arrive at the five non-DCI letters in $z$ (see section $3$).

For other kinematics beyond seven points, cluster variables also successfully predict letters for certain classes of Feynman integrals. For instance, the nine cluster variables from the $D_3$ above are enough for penta-box-ladder integrals (with $(8,5)$ A kinematics), to arbitrary loops~\cite{He:2020uxy,He:2021esx}.
The $9$ DCI letters of the $9{+}7$ ${\cal A}$-coordinates (see appendix) read 
\begin{equation}\label{eq:A3alphabet1}
\{u,\ v,\ w,\ 1{-}u,\ 1{-}v,\ 1{-}w,
1{-}u w,\ 1{-}v w,\ 1{-}u{-}v{+}u v w \},
\end{equation}
for which we denote as $W_1, \cdots, W_9$ in section \ref{sec:3}. Using \eqref{1}, they are equivalent to $9$ positive polynomials, $f_1,f_4,f_6$ and : 
\begin{equation}
\{1{+}f_1,\ 1{+}f_4,\ 1{+}f_6, 1{+}f_4{+}f_1f_4,\ 1{+}f_1{+}f_1f_6,\ 1{+}f_6{+}f_4f_6\}. 
\end{equation}
Generally, most cluster algebras we encounter when $n\geq8$ are infinite type (with some exceptions such as $D_5$ and $D_6$ mentioned above). On the other hand, general conformal integrals with $n\geq 8$, even when they evaluate to MPLs, may contain symbol letters that are algebraic (non-rational) functions of Pl\"ucker coordinates, which certainly go beyond cluster variables. At least for eight points, there is considerable evidence that these algebraic letters are still controlled by affine cluster algebras, and in particular we may extract them from limit rays of tropicalization/Minkowski sum~\cite{Drummond:2019cxm,Henke:2021avn,Drummond:2019qjk,Henke:2019hve}. For eight points, such algebraic letters found so far are associated with two cylclically related square roots of four-mass kinematics, or $A_{1,1}$, which is the ``smallest" in this chain of sub-algebras:
\begin{equation}E_{7}^{(1,1)}\supset E_{6,1} \supset D_{4,1} \supset A_{2,1} \supset A_{1,1}\,. \end{equation}
Algebraic letters for these kinematics can be traced to the simplest ones in $A_{1,1}$. The four-mass box with dual points $(x_2, x_4, x_6, x_8)$ is parametrized by 
\begin{equation}{\bf u}=\frac{x_{24}^2x_{68}^2}{x_{26}^2x_{48}^2}={\bf z} \bar {\bf z},\ {\bf v}=\frac{x_{46}^2x_{82}^2}{x_{26}^2x_{48}^2}=(1-{\bf z})(1-\bar {\bf z}),
\end{equation}
where $({\bf z}, \bar{\bf z}, 1-{\bf z}, 1-\bar{\bf z})$ are {\it algebraic letters} containing the square root of $\Delta_{2,4,6,8}:=(1-{\bf v}+{\bf u})^2-4 {\bf u}$; it is natural to write, in addition to their (rational) products $u,v$, their ratios (odd letters)~\footnote{This is the alphabet for a large class of (off-shell) four-point conformal integrals {\it c.f.} \cite{Basso:2017jwq}.} 
\begin{equation}
\left\{\frac{\bf z}{\bar{\bf z}}, \frac{1{-}\bf z}{1{-}\bar{\bf z}}\right\}.
\end{equation} 
Note that in positive kinematics $0<u,v<1$, not only $\Delta_{2,4,6,8}$ but also the odd letters are {\it positive}: this can be easily seen since ${\bf z}, \bar{\bf z}$ are roots of the equation $r^2-(1-{\bf v}+{\bf u}) r+ {\bf u}=0$.

%\begin{equation}  \mathcal S(I_{2468})=\frac{1}{2} \biggl(u\otimes\frac{1-\bar z}{1-z}+v\otimes \frac{z}{\bar z}\biggr),\end{equation}
For higher sub-algebras containing $A_{1,1}$, all algebraic letters found in previous computations can be obtained similarly from limit rays, and they take the form $\frac{r_i-{\bf z}}{r_i-\bar{\bf z}}$ where $r_i$ are rational functions (including $r_i=0,1$). We have two observations: (1). the product $(r_i-{\bf z})(r_i-{\bar {\bf z}})$ always gives a monomial of cluster variables in the corresponding algebra, which guarantees that these letters have definite signs; (2) there are exactly $d$ multiplicatively independent ratios (odd letters): for $A_{2,1}, D_{4,1}, E_{6,1}$ and the sector with $\Delta_{2,4,6,8}$ of $E_7^{(1,1)}$, we find exactly $3, 5, 7$ and $9$ such $r_i$'s that give independent ratios. Similarly all the rational letters can be obtained by truncating these affine cluster algebras by the method of~\cite{Drummond:2019cxm,Henke:2021avn,Drummond:2019qjk,Henke:2019hve}: for the above cases with $d=7,5,3$, by using non-vanishing Pl\"ucker coordinates for tropicalization/Minknowski sum we find $100, 38$ and $9$ rational letters in the truncated sub-algebras. Alternatively we could select those letters that only depend on the kinematics from the truncated $E_{7}^{(1,1)}$ alphabet ({\it e.g.} with $356$ cluster variables~\cite{Henke:2019hve}): these give $93, 33$ and $8$ rational letters for these cases. We emphasize that such ``truncated" alphabets are not unique, but  they already provide highly non-trivial predictions for alphabets (singularities) of corresponding Feynman integrals: {\it e.g.} for $D_{4,1}$ case, such an alphabet with $38$ (or $33$) rational letters and $5$ algebra ones have been used to to bootstrap the double-penta-ladder at least through $L=4$~\cite{He:2021esx}. 

%In the ancillary file {\bf CA.m}, we present definitions of all $\mathcal{X}$-coordinates $f_i$ for the $n=8$ quivers in the table and $n=6,7$ in Fig.~\ref{fig:folding}, in terms of Pl\"ucker coordinates. For each quiver, we also give the corresponding parametrization of momentum twistors $\mathbf{Z}$ using $f_i$'s. Moreover, we also record alphabets for all these cases in terms of $f_i$, including folding to $D=3$ for $n=6,7$. %We present the alphabet for $D_4$ kinematics in $f_i$s, whose corresponding quiver can be obtained from FIG\ref{fig:folding} by freezing $f_2$ and $f_5$. 
%For those algebras of infinite type with $n=8$, their alphabets contain both rational and algebraic functions of $f_i$, and we select rational letters from the $356$ letters of $G(4,8)$. 

We record all these (truncated) alphabets in the file {\bf CA.m}. %In the end of the file, we also provide parametrization of $\{W_i\}_{i=1\cdots 49}$ from \cite{Abreu:2020jxa} in terms of $f_i$ of $(8,6)$ C.

\section{Bootstrapping three-loop wheel: new algebraic letters from $D_3$}
\label{sec:3}

Due to its success in bootstrap computations, the cluster algebra $G(4,n)$ and its associated alphabet are believed to fully describe all symbol letters that appear in scattering amplitudes. However, for Feynman integrals in the $\mathcal{N}{=}4$ sYM theory, this is {\it not} always the case. The counterexamples appear when the integrals are beyond ladder topology, for instance three-loop wheel integrals in \cite{Bourjaily:2019hmc}. In fact it is intriguing that the nine-point wheel with $D_6$ kinematics is expected to evaluate to functions as iterative integral over K3 surface~\cite{Bourjaily:2019hmc}! In this work we only consider one of its eight-point wheel sub-topology, obtained from the following soft limit:
\begin{center}
    \begin{tikzpicture}[baseline={([yshift=-.5ex]current bounding box.center)},scale=0.12]
        \draw[black,thick] (0,0)--(0,4)--(3.46,6)--(6.93,4)--(6.93,0)--(3.46,-2)--cycle;
        \draw[black,thick] (6.93,4)--(8.66,5);
        \draw[black,thick] (0,4)--(-1.73,5);
        \draw[black,thick] (3.46,-2)--(3.46,-4);
        \draw[black,thick] (0,4)--(3.46,2)--(3.46,-2);
        \draw[black,thick] (3.46,2)--(6.93,4);
        \filldraw[black] (-1.73,5) node[anchor=south east] {{$1$}};
        \filldraw[black] (2.26,7.73) node[anchor=south] {{$2$}};
        \filldraw[black] (4.66,7.73) node[anchor=south] {{$3$}};
        \filldraw[black] (8.66,5) node[anchor=south west] {{$4$}};
        \filldraw[black] (8.93,0) node[anchor=west] {{$5$}};
        \filldraw[black] (7.93,-1.73) node[anchor=north] {{$6$}};
        \filldraw[black] (3.46,-4) node[anchor=north] {{$7$}};
        \filldraw[black] (-1,-1.73) node[anchor=north] {{$8$}};
        \filldraw[black] (-2,0) node[anchor=east] {{$9$}};
        \draw[black,thick] (2.46,7.73)--(3.46,6)--(4.46,7.73);
        \draw[black,thick] (-2,0)--(0,0)--(-1,-1.73);
        \draw[black,thick] (8.93,0)--(6.93,0)--(7.93,-1.73);
        %\filldraw[black] (3.46,2) node[anchor=center] {{6D}};
    \end{tikzpicture} 
$\xrightarrow{\text{soft limit}}$
        \begin{tikzpicture}[baseline={([yshift=-.5ex]current bounding box.center)},scale=0.12]
        \draw[black,thick] (0,0)--(5,0)--(6.55,4.76)--(2.50,7.69)--(-1.55,4.76)--cycle;
        \draw[black,thick] (1.5,9.43)--(2.5,7.69)--(3.5,9.43);
        \draw[black,thick] (-0.21,-1.99)--(0,0)--(-1.83,-0.81);
        \draw[black,thick] (6.83,-0.81)--(5,0)--(5.21,-1.99);
        \draw[black,thick] (2.5,0)--(2.5,3.5)--(6.55,4.76);
        \draw[black,thick] (2.5,3.5)--(-1.55,4.76);
         \draw[black,thick] (6.55,4.76)--(8.45,6.37);
        \draw[black,thick] (-3.55,6.37)--(-1.55,4.76);
        \filldraw[black] (1.5,9.43) node[anchor=south] {{$2$}};
\filldraw[black] (3.5,9.43) node[anchor=south] {{$3$}};
\filldraw[black] (8.45,6.37) node[anchor=west] {{$4$}};
\filldraw[black] (6.83,-0.81) node[anchor=west] {{$5$}};
\filldraw[black] (5.21,-1.99) node[anchor=north] {{$6$}};
\filldraw[black] (-0.21,-1.99) node[anchor=north] {{$7$}};
\filldraw[black] (-1.83,-0.81) node[anchor=east] {{$8$}};
\filldraw[black] (-3.55,6.37) node[anchor=east] {{$1$}};
    \end{tikzpicture}
\end{center}
Although it has simple $D_3$ kinematics, the corresponding Feynman integrals contain new {\it square root} and {\it algebraic letters}. By computing its (composite) leading singularity \cite{Arkani-hamed:2010pyv}, we find ${\rm LS}=\frac{u v w}{\sqrt{\Delta_{D_3}}}$ with 
\begin{equation}
\Delta_{D_3}:=(u+v)^2-4 u v w
\end{equation}
Direct integrations become difficult due to the presence of this new square root.%, for which it suffices to compute the  leading singularity and we find ${\rm LS}=\frac{u v w}{\sqrt{\Delta_{D_3}}}$ with $\Delta_{D_3}:=(u+v)^2-4 u v w$%~\footnote{There is an obstruction for computing the leading singularity of $n=9$ wheel, which is related to the expectation that it goes beyond MPLs; we find an irreducible degree-$8$ polynomial of two parameters under a square root, and it would be interesting to see if the latter is related to $D_6$ cluster algebra!}

Similar to the four-mass box integral, this wheel integral must contain algebraic letters which are odd under $\sqrt{\Delta_{D_3}} \to -\sqrt{\Delta_{D_3}}$; we do not know how to obtain this square root from {\it e.g.} tropicalization of $G(4,8)$. It is natural to ask what algebraic letters can appear, and we show that they are essentially dictated by the $D_3$ cluster algebra!

Note that $\Delta_{D_3}$ naturally appears in the following quadratic equation, $Q(r):=r^2- (u+v) r + u v w=0$, where the two roots $z, \bar{z}$, satisfy $z+\bar{z}=u+v$ and $z \bar{z}=u v w$. Similar to the four-mass box kinematics, we find that in positive kinematics, not only do we have real roots, {\it i.e.} $\Delta_{D_3}>0$, but again $z/\bar{z}$ and $(1-z)/(1-\bar{z})$ are positive since $Q(0)=u v w$ and $Q(1)=W_9$ are monomials of cluster variables! Now we can look for other rational $r_i$'s such that $(r_i{-}z)(r_i{-}\bar{z})$ becomes such a monomial with definite sign. The solutions we find are
\begin{equation}
    \left\{0,1,u,v,\frac{w(1{-}u)}{1{-}u w}:=r_1, \frac{w(1{-}v)}{1{-}v w}:=r_2\right\}.
\end{equation}
%$r=u,v$  and some more complicated ones denoted as $r_1{=}w (1{-}u)/(1{-}u w)$ and $r_2{=}r_1|_{u\to v}$; 
Remarkably, the corresponding odd letters satisfy
\begin{equation} \frac{u{-}z}{u{-}\bar{z}}\frac{v{-}z}{v{-}\bar{z}}=1,\quad \frac{r_1{-}z}{r_1{-}\bar{z}}= \frac{z}{\bar{z}}\frac{u{-}z}{u{-}\bar{z}}\frac{1{-}\bar{z}}{1{-}z}
\end{equation}
also for $r_2$ in the second relation with $u$ and $v$ swapped. We find exactly three multiplicatively-independent algebraic letters (in accordance with the number of degrees of freedom), and we choose them to be 
\begin{equation}
    \{L_i\}_{i=1,2,3}:=\left\{\frac{r_1-z}{r_1-\bar{z}},\frac{r_2-z}{r_2-\bar{z}},\frac{z}{\bar{z}}\right\}
\end{equation}

Based on these three odd letters together with the nine even letters \eqref{eq:A3alphabet1}, we are ready to bootstrap the symbol of eight-point wheel. After imposing physical first entries being $u,v,w$, we construct integrable symbols with nine rational letters and three algebraic ones: at weight six for three loops, we find $4773$ such integrable symbols. Next, note the integral is symmetric in $u,v$, and each term of its symbol must contains odd number of $L_i$'s; with these the number of integrable symbols is reduced to $737$. Furthermore, we can impose two boundary conditions: with either $w\to 1$  or $W_9\to 0$, $\Delta_{D_3}$ becomes a perfect square, and it is straightforward to evaluate the integral in these limits (both $A_2$ functions)~\cite{Bourjaily:2019hmc}. These boundary conditions reduces the number down to $278$.

Finally, we impose the Steinmann relations \cite{Steinmann1960a,Steinmann1960b, Caron-Huot:2016owq}%(see also \cite{Bourjaily:2020wvq,Dixon:2020bbt,Benincasa:2020aoj,He:2021mme})
, which turns out to be the most constraining: for each pair of incompatible channels, we ask the double discontinuity to vanish. Surprisingly, the constraint from one pair of channels, {\it e.g.} $\langle1267\rangle$ and $\langle3481\rangle$, already fixes all coefficients and we arrive at a unique symbol! In particular, it excludes all terms which contain algebraic letters more than once, and in fact $L_i$'s are constrained to appear only in the last entry:
\begin{equation}
    \mathcal{S}(I_{n=8}^{\rm wheel})= \sum_{i=1}^3 \mathcal{S}(F_i(\{W\}))\otimes L_i
\end{equation}
where $F_i$ for $i=1,2,3$ are three weight-five $D_3$  functions! We record the symbol for wheel integral, together with its two special limits in the ancillary file \textbf{wheel.m}. 

Steinmann relations for any other pair of channels and all extended Steinmann relations (beyond the first two entries) become very strong cross checks of our result, and it indeed obeys all these constraints. Nicely, we find extended Steinmann relations between $W_i$'s are equivalent to {\it cluster adjacency conditions} of $D_3$ in terms of $\mathcal{A}$-coordinates; {\it i.e.}  any two mutable cluster variables cannot appear adjacent in the symbol if they do not show up in the same cluster (one of the $14$ for $D_3$)~\footnote{Frozen variables can be adjacent to any one, and these conditions have also been checked for penta-box ladders through four loops~\cite{He:2021mme}}.

\section{Folding to kinematics in three dimensions}
In this section, we present a remarkable link between dimension reduction to three dimensions and folding of cluster algebras for kinematics.

The dimension reduction to three dimensions can be achieved by setting {\it Gram determinants} to zero: for $x^\mu$ to lie in three dimensions ($X^I$ to lie in five dimensions), any six $X$'s must have linearly dependent, which means the $6\times 6$ matrix made of $X_i \cdot X_j$ must have vanishing determinant~\footnote{These Gram determinant conditions define the parity-invariant subspace~\cite{Dixon:2021tdw}, since in spinor variables going to three dimensions amounts to identify $\tilde\lambda$ with $\lambda$ which are related to each other by parity. They are equivalent to conditions on momentum twistors given in~\cite{Elvang:2014fja}.}. Since $X_i \cdot X_j=x_{i,j}^2 \propto \langle i{-}1 i j{-}1 j\rangle$, we ask that for any $6$ dual points, $x_{i_1}, x_{i_2}, \cdots, x_{i_6}$, the matrix with entries $G_{a,b}:=\langle i_a{-}1 i_a i_b{-}1 i_b \rangle$ (for $1\leq a, b\leq 6$) must have $\det G=0$. For massless kinematics, exactly $n-5$ conditions are independent, which thus gives $2(n{-}5)$-dim DCI kinematics in three dimensions as expected.  The key observation is that, for positive kinematics, such Gram-determinant conditions become simple conditions on cluster variables which realize three-dimensional kinematics as folded cluster algebras!

\begin{figure}
    \centering
    \includegraphics[scale=1]{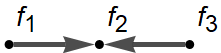}
    \includegraphics[scale=0.6]{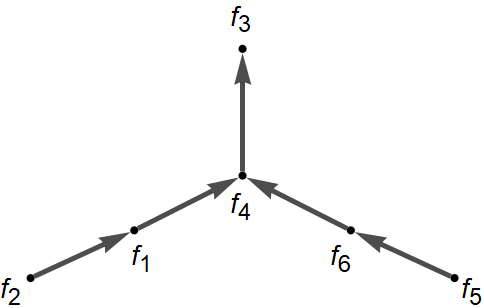}
    \caption{Quivers for $A_3$ and $E_6$ making folding explicit.}
    \label{fig:folding}
\end{figure}

Let us look at the simplest six-point massless kinematics, where the only Gram determinant is proportional to \begin{equation}
\Delta{=}\sqrt{(1{-}u_1{-}u_2{-}u_3)^2{-}4u_1u_2u_3},\ \ \ \ u_1=\frac{x_{1,3}^2x_{4,6}^2}{x_{1,4}^2x_{3,6}^2}\ \ \ \  \&\  cyc.
\end{equation}
It is known~\cite{Chicherin:2020umh} that with $\Delta=0$, the $A_3$ cluster algebra is folded into a $C_2$: only the first six of the nine letters
\begin{equation}
\{u_i, 1-u_i, y_i\}_{i=1,2,3},\ \ y_1{=}\frac{1{+}u_1{-}u_2{-}u_3{-}\Delta}{1{+}u_1{-}u_2{-}u_3{+}\Delta}
\end{equation} 
survive since $y_i\to 1$. To see the folding explicitly, we choose a quiver as in Fig \ref{fig:folding}), where 
\begin{equation}
f_1{=}\frac{\langle1234\rangle\langle1256\rangle}{\langle1236\rangle\langle1245\rangle},\ f_2{=}\frac{\langle1246\rangle\langle1345\rangle}{\langle1234\rangle\langle1456\rangle},\ f_3{=}\frac{\langle1456\rangle\langle2345\rangle}{\langle1245\rangle\langle3456\rangle}.
\end{equation} 
Very nicely, $\Delta=0$ is equivalent to $f_1=f_3$! %this condition reduces the $9$ letters of $A_3$ to the $6$ of $C_2$:\begin{align}\{f_1,f_2,1{+}f_1,1{+}f_2,1{+}f_1{+}f_1f_2,1{+}2f_1{+}f_1^2{+}f_1^2f_2\}.\end{align}

Similarly, for the seven-point massless case there are two independent Gram-determinant conditions, which reduce the $42$ letters of $E_6$ down to $28$ letters of an $F_4$ cluster algebra! In terms of $u_1=x^2_{2,4}x^2_{5,1}/x^2_{2,5}x^2_{4,1}$ and $u_i=u_1(Z_j\to Z_{j{+}i{-}1})$, there are $28$ parity-even letters that are rational functions, and the remaining parity-odd ones all become one when going to three dimensions, resulting in 
\begin{equation}
\{u_i,1{-}u_i,1{-}u_i u_{i{+}3},1{-}u_i u_{i{+}3}{-}u_{i{+}1}u_{i{+}5}\}_{i=1,\cdots, 7}\,.
\end{equation}
It is highly non-trivial that these $28$ letters (with constraints so only $d=4$) form an $F_4$ cluster algebra, but it can be made manifest with the quiver in Fig.~\ref{fig:folding}: the definition of $f_i$'s can be found in {\bf CA.m}, and the two conditions become simply $f_2=f_5$, $f_1=f_6$. Despite the cluster algebras become infinite type, we have found exactly the same folding for eight and nine points with three dimensions, where the quivers can be chosen such that $n{-}5$ conditions are simply $f_i=f_{3(n{-}5)-i}$ for $i=1,\cdots, n{-}5$.

For general kinematics with $m\geq 6$, we expect $m{-}5$ independent conditions and the degrees of freedom in three dimensions becomes $n{+}m{-}10$. Again remarkably in all cases we have studied this corresponds to folding of cluster algebras. For example, the $D_d$ cluster algebras (with $d=3,4,5,6$) in three dimensions become $B_{d{-}1}$, and even the affine types $E_{6,1}$ and $D_{4,1}$ become $F_{4,1}$ and $B_{3,1}$ respectively! 

We can apply these results to conformal integrals in three dimensions such as those naturally appeared for ABJM amplitudes. The only data we have found in the literature is the two-loop six-point amplitude as well as various finite DCI integrals contained in it~\cite{Caron-Huot:2012sos}. From these weight-$2$ functions, we find the $6$ letters of $C_2$, $\{u_i, 1-u_i\}_{i=1,2,3}$ as well as letters that are odd under $\mathbb{Z}_2$ little group, $\chi_i:=(u_i + \sqrt{u_i-1})/(u_i-\sqrt{u_i-1})$. Note that $\chi_i$ still has singularities at those of $C_2$ alphabet ($u_i=0,1$) and it is a phase factor with positive kinematics. We conjecture that these $6+3$ letters may be all we need for higher-loop amplitudes in ABJM. Although seven-point ABJM amplitude vanishes, we can compute a large class of finite DCI two-loop integrals~\cite{pricom}: at weight two we find exactly the first $21$ of the $28$ letters in $F_4$ (we expect the remaining seven at higher weights), and seven odd letters $\chi_i$ for $i=1,\cdots, 7$ as well! It is an interesting question if these $28+7$ letters are sufficient for higher-loop integrals. There is evidence that starting eight points, even two-loop ABJM amplitudes contain integrals that evaluate to elliptic functions~\cite{pricom}, which are beyond the scope of this letter. 

\section{Applications for non-conformal integrals}
Before ending, we apply some of our results to non-conformal Feynman integrals, following the method of~\cite{Chicherin:2020umh}. The key observation is that by sending a generic dual point (not light-like separated from any other points, or one that is between two massive corners) to {\it infinity}, conformal symmetry is broken and we obtain a non-DCI kinematics from such a DCI one~\footnote{A well-known example is that the non-DCI kinematics of a triangle (parametrized by two ratios of three masses) can be obtained by sending a point of four-mass box kinematics to infinity.}. 

As a warm-up exercise, by sending $x_6 \to \infty$ for the DCI kinematics $(7,5)$ B, one arrives at a non-DCI kinematics with three massless leg and a massive one (dual points $x_1, x_2, x_3, x_4$). Define \begin{equation}
s=x_{1,3}^2,\ t=x_{2,4}^2,\ m^2=x_{1,4}^2,
\end{equation}
we see that the five letters of $A_2$ cluster algebra is mapped to 
\begin{equation}
\left\{z_1:=\frac{s}{m^2}, z_2:=\frac{t}{m^2}, 1{-}z_1,\  1{-}z_2,\ \  1{-}z_1{-}z_2\right\}.
\end{equation}
We have restricted the massive corner to lie between leg $3$ and $1$, but if we allow other positions, we have an additional letter $(s+t)/m^2=z_1+z_2$, which nicely recovers the $C_2$ alphabet for four-point Feynman integrals with a massive leg. This is the same $C_2$ for three-point form factor in ${\cal N}=4$ sYM through eight loops (although obtained as folding of $A_3$ in~\cite{Chicherin:2020umh})!

Now we move to non-DCI limit of $(8,5)$A kinematics: by sending the dual point $x_7$ to infinity, we get non-DCI two-mass-easy box kinematics (dual points $x_1, x_2, x_4, x_5$).
\begin{figure}[t]
\centering
    \begin{tikzpicture}[scale=0.2]
        \draw[black,thick] (0,0)--(5,0)--(6.55,4.76)--(2.50,7.69)--(-1.55,4.76)--cycle;
        \draw[black,thick] (1.5,9.43)--(2.5,7.69)--(3.5,9.43);
        \draw[black,thick] (-0.21,-1.99)--(0,0)--(-1.83,-0.81);
        \draw[black,thick] (6.83,-0.81)--(5,0)--(5.21,-1.99);
         \draw[black,thick] (6.55,4.76)--(8.45,6.37);
        \draw[black,thick] (-3.55,6.37)--(-1.55,4.76);
        \filldraw[black] (1.5,9.43) node[anchor=south] {{$2$}};
\filldraw[black] (3.5,9.43) node[anchor=south] {{$3$}};
\filldraw[black] (8.45,6.37) node[anchor=west] {{$4$}};
\filldraw[black] (6.83,-0.81) node[anchor=west] {{$5$}};
\filldraw[black] (5.21,-1.99) node[anchor=north] {{$6$}};
\filldraw[black] (-0.21,-1.99) node[anchor=north] {{$7$}};
\filldraw[black] (-1.83,-0.81) node[anchor=east] {{$8$}};
\filldraw[black] (-3.55,6.37) node[anchor=east] {{$1$}};
\filldraw[black] (10,4) node[anchor=west] {{$\longrightarrow$}};
    \end{tikzpicture}
    \begin{tikzpicture}[scale=0.2]
                \draw[black,thick] (0,5)--(-5,5)--(-5,0)--(0,0)--cycle;
                \draw[black,thick] (0,5)--(0.52,6.93);
                \draw[black,thick] (1.93,-0.52)--(0,0)--(0.52,-1.93);
                \draw[black,thick] (-6.93,5.52)--(-5,5)--(-5.52,6.93);
                \draw[black,thick] (-5,0)--(-5.52,-1.93);
                \filldraw[black] (0.52,6.93) node[anchor=south west] {{$4(P_2)$}};
                \filldraw[black] (1.93,-1) node[anchor=west] {{$3$}};
                \filldraw[black] (0.52,-1.93) node[anchor=north] {{$2$}};
                \filldraw[black] (1,-2.5) node[anchor=north west] {{$(P_1)$}};
                \filldraw[black] (-6.93,6) node[anchor=east] {{$8$}};
                \filldraw[black] (-5.52,6.93) node[anchor=south] {{$5$}};
                 \filldraw[black] (-6,7.5) node[anchor=south east] {{$(P_3)$}};
                \filldraw[black] (-5.52,-1.93) node[anchor=north east] {{$(P_4)1$}};
            \end{tikzpicture}   
        \caption{$(8,5)$ A kinematics (left) and its non-DCI limit (right)}
            \label{fig:pentagon}
\end{figure}
%Regarding any massive corners as massive legs, kinematics variables $\{s,t,m_1^2,m_3^2\}$ then have momentum twistors parametrization as\begin{align}  s&=\frac{\langle1245\rangle}{\langle1267\rangle\langle4567\rangle},\ t=\frac{\langle3481\rangle}{\langle3467\rangle\langle8167\rangle},\ \nonumber\\   m_1^2&=\frac{\langle1234\rangle}{\langle1267\rangle\langle3467\rangle},\    m_3^2=\frac{\langle4581\rangle}{\langle8167\rangle\langle4567\rangle}\end{align}
In this case $s=x_{2,5}^2$, $t=x_{1,4}^2$, $m_1^2=x_{2,4}^2$, $m_3^2=x_{1,5}^2$, and we find that the cross-ratios are mapped to $u=s/{m_3^2},\ v=t/{m_3^2},\ w=(m_1^2 m_3^2)/(s t)$. Therefore, the $9$ cluster variables of $D_3$ become $9$ ratios of 
\begin{align}
    \{s,\ t,\ m_1^2,\ m_3^2,\ s{-}m_1^2,\ t{-}m_1^2,\ s{-}m_3^2,\ t{-}m_3^2, 
    s t{-}m_1^2m_3^2,\ s{+}t{-}m_1^2{-}m_3^2\}
\end{align}
which exactly reproduces the alphabet for (the family of) one-loop two-mass-easy box integral~\cite{Chicherin:2020umh} (note that we recover the $A_2$ alphabet above with $m_3^2\to 0$).  

It was known that these nine letters are insufficient for two-loop integrals with two-mass-easy kinematics; in particular, three additional letters containing a new square root $\Delta^{nc}$ were found: in the notation of~\cite{Abreu:2020jxa} they are $\{W_{35},W_{36},W_{39}\}$ for the two-mass-easy sector~\footnote{Note that these $3$ odd letters also show up when we degenerate the alphabet of (the family) of two-loop three-mass double box integrals \cite{Dlapa:2021qsl}.}. As we learn from our wheel example, for $D_3$ kinematics we need to supplement the $9$ rational letters with three algebraic ones. Remarkably, we find that $\{L_1, L_2, L_3\}$ in the limit spans exactly the same space of $\{W_{35},W_{36},W_{39}\}$! In particular, we find $\Delta_{D_3} \propto (s+t)^2-4 m_1^2 m_3^2=\Delta^{nc}$ under the map. This further supports the robustness of our generalized $D_3$ alphabet: in the non-DCI limit, the $9+3$ letters of $D_3$ kinematics precisely coincide with those for two-mass-easy integrals at two loops!

Finally, the non-DCI kinematics with four massless legs and a massive one can be obtained by sending $x_7 \to \infty$ of our $(8,6)$C kinematics~\cite{Chicherin:2020umh}, 
%However, there are situations that cluster variables are  not enough to predict even the rational letters, says non-DCI one-mass pentagon, which is the non-DCI limit $(67)\to\infty$ of the $(8,6)$ C kinematics:
where $P_1^2=m_1^2$ for the massive corner. We have the relations $u={s_{12}}/{m_1^2}$,\ $v={s_{51}}/{m_1^2}$,\ $w=(m_1^2s_{34})/(s_{12}s_{51})$,\ $u^\prime={s_{23}}/{m_1^2}$ and $v^\prime={s_{45}}/{m_1^2}$, and it is interesting to compare with the $49$ relevant letters of~\cite{Abreu:2020jxa} (see also~\cite{Guo:2021bym}). First we note that $\Delta_{D_3}$, $\Delta_{2,4,6,8}$ and the Gram determinant (proportional to $f_2-f_7$) corresponds to $\Delta^{nc}$, $W_{48}$ and $W_{49}$ respectively; for $\{W_i\}_{i=1\cdots 47}$, we find that all algebraic letters are reproduced and so do most of the rational letters (depending on precise truncation). However, some rational letters %{\it e.g.} $\{W_{25}, W_{26}, W_{31},W_{32},W_{41},W_{42}\}$
contain factors that are not cluster variables, {\it e.g.} $f_2{-}f_6$, $f_6{-}f_7$, which in some sense are analog of the Gram determinant. We record the parametrization of $W_{i=1,\cdots 49}$ using $D_{4,1}$ cluster variables in the file {\bf CA.m}. 
%\begin{align}&\{f_2{-}f_6,\ f_6{-}f_7,\ 1{-}f_1f_4f_7,\ 1{-}f_1f_2f_4,\     &f_1 f_2f_4^2 f_7 {+}f_1 f_2 f_4{+}f_1 f_4 f_7{+}f_1 f_2 f_4 f_7{-}1,\\    &f_1 f_6{+}f_1 f_2 f_6{+}f_1 f_6 f_7{+}f_6{-}f_1 f_2 f_7\}\,.\end{align}

\begin{figure}[t]
\centering
    \begin{tikzpicture}[scale=0.2]
                \draw[black,thick] (0,0)--(4,0)--(6,3.46)--(4,6.93)--(0,6.93)--(-2,3.46)--cycle;
                \draw[black,thick] (0,6.93)--(-1,8.66);
                \draw[black,thick] (6,7.22)--(4,6.93)--(4.5,8.66);
                \draw[black,thick] (7.74,2.46)--(6,3.46)--(7.74,4.46);
                \draw[black,thick] (4,0)--(5,-1.73);
                \draw[black,thick] (0,0)--(-1,-1.73);
                \draw[black,thick] (-2,3.46)--(-4,3.46);
                \filldraw[black] (-1,8.66) node[anchor=south east] {{4}};
                \filldraw[black] (4.5,8.66) node[anchor=south west] {{5}};
                \filldraw[black] (6,7.22) node[anchor=south west] {{6}};
                \filldraw[black] (7.74,4.46) node[anchor=west] {{7}};
                \filldraw[black] (7.74,2.46) node[anchor=west] {{8}};
                \filldraw[black] (5,-1.73) node[anchor=north west] {{1}};
                \filldraw[black] (0,-1.73) node[anchor=north east] {{2}};
                \filldraw[black] (-4,3.46) node[anchor=east] {{3}};
            \end{tikzpicture}
             \begin{tikzpicture}[scale=0.2]
        \draw[black,thick] (0,0)--(5,0)--(6.55,4.76)--(2.50,7.69)--(-1.55,4.76)--cycle;
        \draw[black,thick] (1.5,9.43)--(2.5,7.69)--(3.5,9.43);
        \draw[black,thick] (0,0)--(-1.83,-0.81);
        \draw[black,thick] (6.83,-0.81)--(5,0);
         \draw[black,thick] (6.55,4.76)--(8.45,6.37);
        \draw[black,thick] (-3.55,6.37)--(-1.55,4.76);
        \filldraw[black] (1.5,9.43) node[anchor=south] {{$5$}};
\filldraw[black] (3.5,9.43) node[anchor=south] {{$8$}};
\filldraw[black] (8.45,6.37) node[anchor=west] {{$1$}};
\filldraw[black] (6.83,-0.81) node[anchor=west] {{$2$}};
\filldraw[black] (-1.83,-0.81) node[anchor=east] {{$3$}};
\filldraw[black] (-3.55,6.37) node[anchor=east] {{$4$}};
\filldraw[black] (2.5,11.2) node[anchor=south] {{$(P_1)$}};
\filldraw[black] (10.45,8.37) node[anchor=west] {{$(P_2)$}};
\filldraw[black] (8.83,-2.81) node[anchor=west] {{$(P_3)$}};
\filldraw[black] (-3.83,-2.81) node[anchor=east] {{$(P_4)$}};
\filldraw[black] (-5.55,8.37) node[anchor=east] {{$(P_5)$}};
\filldraw[black] (-10,4) node[anchor=east] {{$\longrightarrow$}};
    \end{tikzpicture}
\caption{$(8,6)$ C kinematics (left) and its non-DCI limit (right)}
            \label{fig:hex}
\end{figure}

\section{Conclusions}
In this work, we have proposed parametrizing planar kinematics of conformal integrals as sub-algebras of $G(4,n)$ cluster algebras, and argued that cluster variables and their algebraic generalizations know about singularities of DCI Feynman integrals; in addition to evidence for various ladder integrals, we have demonstrated the power of cluster algebras and adjacency by bootstrapping the three-loop wheel from a generalized $D_3$ alphabet. We have also outlined applications to more general, non-DCI Feynman integrals, as well as to those in three dimensions via folding. There seems to be certain ``universality" of cluster algebras in a wide range of integrals and physical quantities.

There are numerous questions raised by our investigations. An important mathematical question is if one could show, for any kinematics, there always exists one (and only one) sub-algebra, which would also make our ``freezing" algorithm more systematic. We have seen that DCI Feynman integrals, even when evaluated to MPLs, can have more complicated ({\it e.g.} algebraic) singularities than naively expected from the alphabet of cluster algebras, but they are still strongly constrained by the latter, as illustrated by the $D_3$ wheel example. It would be highly desirable to see how universal is this phenomenon and understand its origin. Relatedly, it is tempting to ask if singularities of Feynman integrals beyond MPLs (such as the $D_6$ nine-point wheel~\cite{Bourjaily:2019hmc}, or ten-point double box integral~\cite{Bourjaily:2017bsb, Kristensson:2021ani}) may also be constrained by corresponding cluster algebras? So far it seem totally miraculous that cluster algebras (and cluster adjacency) can be applied to form factors, amplitudes/integrals in three dimensions, and non-DCI Feynman integrals {\it etc.}. Is it possible to have more systematic understanding, perhaps through the connection of cluster algebras with canonical differential equations satisfied by Feynman integrals of a given kinematics~\cite{Henn:2013pwa,Henn:2014qga}?

\section*{Acknowledgments}
We thank Nima Arkani-Hamed, James Drummond, Yu-tin Huang, Chia-kai Kuo, Georgios Papathanasiou and Yichao Tang for inspiring discussions. S.H. thanks organizers and participants of the workshop ``Scattering Amplitudes, Cluster Algebras, and Positive Geometries" where part of the results were reported. This research is supported in part by National Natural Science Foundation of China under Grant No. 11935013,11947301,12047502,12047503.

%\appendix

%\section{Details of the quivers and alphabets for kinematics}
%In this appendix we provide some details in finding the sub-quivers for planar kinematics as well as their alphabet. Let us first list possible $6$-point and $7$-point kinematics  except for the fully massless ones in FIG.\ref{fig:67ptkinemaics}. As discussed in section I and shown in Fig.~\ref{fig:6pt}, we find an $A_1$ sub-algebra for $(6,5)$: there are $2$ unfrozen variables $\langle1346\rangle$, $\langle1245\rangle$ and $4$ frozen ones $\langle1234\rangle$,$\langle1456\rangle$,$\langle1345\rangle$ and $\langle1246\rangle$.

%\begin{figure}
%        \centering
%        \subfigure[$n{=}6$] {\includegraphics[scale=0.65]{6ptAcoordinate.pdf}}
%        \subfigure[$n=7$]  
%        {\includegraphics[scale=0.5]{7ptAcoordinate.pdf}}
       %{$n=6$: blue nodes denote the original frozen variables, green and red nodes are non-frozen nodes of $A_3$ for the Grassmannian $G(4,6)$, and red node ($A_1$) describes the $(6,5)$ kinematics;}

        %{blue nodes denote the original frozen variables, green nodes denote the further frozen node to get $D_4$ for $(7,6)$; red sub-quiver is the $A_2$ corresponds to $(7,5)$ A}
      
   % \end{figure}

\bibliographystyle{utphys}
\bibliography{letter}

@PREAMBLE{
 "\providecommand{\noopsort}[1]{}" 
 # "\providecommand{\singleletter}[1]{#1}%" 
}

@article{pricom,
  author = {Huang, Yu-tin and Kuo, Chia-Kai},
  journal   = "{private communications}"
}

@article{Caron-Huot:2012sos,
    author = "Caron-Huot, S. and Huang, Yu-tin",
    title = "{The two-loop six-point amplitude in ABJM theory}",
    eprint = "1210.4226",
    archivePrefix = "arXiv",
    primaryClass = "hep-th",
    reportNumber = "MCTP-12-24",
    doi = "10.1007/JHEP03(2013)075",
    journal = "JHEP",
    volume = "03",
    pages = "075",
    year = "2013"
}

@article{Arkani-Hamed:2010pyv,
    author = "Arkani-Hamed, Nima and Bourjaily, Jacob L. and Cachazo, Freddy and Trnka, Jaroslav",
    title = "{Local Integrals for Planar Scattering Amplitudes}",
    eprint = "1012.6032",
    archivePrefix = "arXiv",
    primaryClass = "hep-th",
    doi = "10.1007/JHEP06(2012)125",
    journal = "JHEP",
    volume = "06",
    pages = "125",
    year = "2012"
}

@article{Basso:2017jwq,
    author = "Basso, Benjamin and Dixon, Lance J.",
    title = "{Gluing Ladder Feynman Diagrams into Fishnets}",
    eprint = "1705.03545",
    archivePrefix = "arXiv",
    primaryClass = "hep-th",
    reportNumber = "SLAC-PUB-16967",
    doi = "10.1103/PhysRevLett.119.071601",
    journal = "Phys. Rev. Lett.",
    volume = "119",
    number = "7",
    pages = "071601",
    year = "2017"
}

@article{He:2021mme,
    author = "He, Song and Li, Zhenjie and Yang, Qinglin",
    title = "{Comments on all-loop constraints for scattering amplitudes and Feynman integrals}",
    eprint = "2108.07959",
    archivePrefix = "arXiv",
    primaryClass = "hep-th",
    month = "8",
    year = "2021"
}

@article{Bourjaily:2019hmc,
    author = "Bourjaily, Jacob L. and McLeod, Andrew J. and Vergu, Cristian and Volk, Matthias and Von Hippel, Matt and Wilhelm, Matthias",
    title = "{Embedding Feynman Integral (Calabi-Yau) Geometries in Weighted Projective Space}",
    eprint = "1910.01534",
    archivePrefix = "arXiv",
    primaryClass = "hep-th",
    doi = "10.1007/JHEP01(2020)078",
    journal = "JHEP",
    volume = "01",
    pages = "078",
    year = "2020"
}

@article{Dixon:2021tdw,
    author = "Dixon, Lance J. and Gurdogan, Omer and McLeod, Andrew J. and Wilhelm, Matthias",
    title = "{Folding Amplitudes into Form Factors: An Antipodal Duality}",
    eprint = "2112.06243",
    archivePrefix = "arXiv",
    primaryClass = "hep-th",
    reportNumber = "SLAC-PUB-17637",
    month = "12",
    year = "2021"
}

@article{Guo:2021bym,
    author = "Guo, Yuanhong and Wang, Lei and Yang, Gang",
    title = "{Bootstrapping a Two-Loop Four-Point Form Factor}",
    eprint = "2106.01374",
    archivePrefix = "arXiv",
    primaryClass = "hep-th",
    doi = "10.1103/PhysRevLett.127.151602",
    journal = "Phys. Rev. Lett.",
    volume = "127",
    number = "15",
    pages = "151602",
    year = "2021"
}

@article{Dixon:2020bbt,
    author = "Dixon, Lance J. and McLeod, Andrew J. and Wilhelm, Matthias",
    title = "{A Three-Point Form Factor Through Five Loops}",
    eprint = "2012.12286",
    archivePrefix = "arXiv",
    primaryClass = "hep-th",
    reportNumber = "SLAC-PUB-17581",
    doi = "10.1007/JHEP04(2021)147",
    journal = "JHEP",
    volume = "04",
    pages = "147",
    year = "2021"
}

@article{Li:2021bwg,
    author = "Li, Zhenjie and Zhang, Chi",
    title = "{The Three-loop MHV Octagon from $\bar{Q}$ equations}",
    eprint = "2110.00350",
    archivePrefix = "arXiv",
    primaryClass = "hep-th",
    month = "10",
    year = "2021"
}

@article{Golden:2021ggj,
    author = "Golden, John and McLeod, Andrew J.",
    title = "{The two-loop remainder function for eight and nine particles}",
    eprint = "2104.14194",
    archivePrefix = "arXiv",
    primaryClass = "hep-th",
    doi = "10.1007/JHEP06(2021)142",
    journal = "JHEP",
    volume = "06",
    pages = "142",
    year = "2021"
}

@article{Kristensson:2021ani,
    author = "Kristensson, Alexander and Wilhelm, Matthias and Zhang, Chi",
    title = "{The elliptic double box and symbology beyond polylogarithms}",
    eprint = "2106.14902",
    archivePrefix = "arXiv",
    primaryClass = "hep-th",
    month = "6",
    year = "2021"
}

@article{Steinmann1960a,
author={Steinmann, O.},
title={{\"U}ber den Zusammenhang zwischen den Wightmanfunktionen und den retardierten Kommutatoren},
journal={Helvetica Physica Acta},
year={1960},
publisher={Birkh{\"a}user},
volume={33},
number={IV},
pages={257},
issn={0018-0238},
doi={10.5169/seals-113076},
url={https://doi.org/10.5169/seals-113076}
}

@article{Steinmann1960b,
author={Steinmann, O.},
title={Wightman-Funktionen und retardierte Kommutatoren. II},
journal={Helvetica Physica Acta},
year={1960},
publisher={Birkh{\"a}user},
volume={33},
number={V},
pages={347},
issn={0018-0238},
doi={10.5169/seals-113079},
url={https://doi.org/10.5169/seals-113079}
}

@article{Abreu:2020jxa,
    author = "Abreu, Samuel and Ita, Harald and Moriello, Francesco and Page, Ben and Tschernow, Wladimir and Zeng, Mao",
    title = "{Two-Loop Integrals for Planar Five-Point One-Mass Processes}",
    eprint = "2005.04195",
    archivePrefix = "arXiv",
    primaryClass = "hep-ph",
    doi = "10.1007/JHEP11(2020)117",
    journal = "JHEP",
    volume = "11",
    pages = "117",
    year = "2020"
}

@article{Arkani-Hamed:2013jha,
  author        = {Arkani-Hamed, Nima and Trnka, Jaroslav},
  title         = {{The Amplituhedron}},
  journal       = {JHEP},
  volume        = {10},
  year          = {2014},
  pages         = {030},
  doi           = {10.1007/JHEP10(2014)030},
  eprint        = {1312.2007},
  archiveprefix = {arXiv},
  primaryclass  = {hep-th},
  slaccitation  = {%%CITATION = ARXIV:1312.2007;%%}
}

@book{Arkani-Hamed:2016byb,
  author        = {Arkani-Hamed, Nima and Bourjaily, Jacob L. and Cachazo, Freddy and Goncharov, Alexander B. and Postnikov, Alexander and Trnka, Jaroslav},
  title         = {{Grassmannian Geometry of Scattering Amplitudes}},
  eprint        = {1212.5605},
  archiveprefix = {arXiv},
  primaryclass  = {hep-th},
  reportnumber  = {PUPT-2435},
  doi           = {10.1017/CBO9781316091548},
  isbn          = {978-1-107-08658-6, 978-1-316-57296-2},
  publisher     = {Cambridge University Press},
  month         = {4},
  year          = {2016}
}

@article{Arkani-Hamed:2019rds,
    author = "Arkani-Hamed, Nima and Lam, Thomas and Spradlin, Marcus",
    title = "{Non-perturbative geometries for planar $ \mathcal{N} $ = 4 SYM amplitudes}",
    eprint = "1912.08222",
    archivePrefix = "arXiv",
    primaryClass = "hep-th",
    doi = "10.1007/JHEP03(2021)065",
    journal = "JHEP",
    volume = "03",
    pages = "065",
    year = "2021"
}

@article{ArkaniHamed:2010gh,
  author        = {Arkani-Hamed, Nima and Bourjaily, Jacob L. and Cachazo,
                        Freddy and Trnka, Jaroslav},
  title         = {{Local Integrals for Planar Scattering Amplitudes}},
  journal       = {JHEP},
  volume        = {06},
  year          = {2012},
  pages         = {125},
  doi           = {10.1007/JHEP06(2012)125},
  eprint        = {1012.6032},
  archiveprefix = {arXiv},
  primaryclass  = {hep-th},
  slaccitation  = {%%CITATION = ARXIV:1012.6032;%%}
}

@article{Bourjaily:2013mma,
  author        = {Bourjaily, Jacob L. and Caron-Huot, Simon and Trnka,
                        Jaroslav},
  title         = {{Dual-Conformal Regularization of Infrared Loop
                        Divergences and the Chiral Box Expansion}},
  journal       = {JHEP},
  volume        = {01},
  year          = {2015},
  pages         = {001},
  doi           = {10.1007/JHEP01(2015)001},
  eprint        = {1303.4734},
  archiveprefix = {arXiv},
  primaryclass  = {hep-th},
  slaccitation  = {%%CITATION = ARXIV:1303.4734;%%}
}

@article{Bourjaily:2017bsb,
  author        = {Bourjaily, Jacob L. and McLeod, Andrew J. and Spradlin,
                        Marcus and von Hippel, Matt and Wilhelm, Matthias},
  title         = {{Elliptic Double-Box Integrals: Massless Scattering
                        Amplitudes beyond Polylogarithms}},
  journal       = {Phys. Rev. Lett.},
  volume        = {120},
  year          = {2018},
  number        = {12},
  pages         = {121603},
  doi           = {10.1103/PhysRevLett.120.121603},
  eprint        = {1712.02785},
  archiveprefix = {arXiv},
  primaryclass  = {hep-th},
  slaccitation  = {%%CITATION = ARXIV:1712.02785;%%}
}

@article{Bourjaily:2018aeq,
  author        = {Bourjaily, Jacob L. and McLeod, Andrew J. and von Hippel,
                        Matt and Wilhelm, Matthias},
  title         = {{Rationalizing Loop Integration}},
  journal       = {JHEP},
  volume        = {08},
  year          = {2018},
  pages         = {184},
  doi           = {10.1007/JHEP08(2018)184},
  eprint        = {1805.10281},
  archiveprefix = {arXiv},
  primaryclass  = {hep-th},
  slaccitation  = {%%CITATION = ARXIV:1805.10281;%%}
}

@article{Caron-Huot:2013vda,
  author        = {Caron-Huot, Simon and He, Song},
  title         = {{Three-loop octagons and $n$-gons in maximally
                        supersymmetric Yang-Mills theory}},
  journal       = {JHEP},
  volume        = {08},
  year          = {2013},
  pages         = {101},
  doi           = {10.1007/JHEP08(2013)101},
  eprint        = {1305.2781},
  archiveprefix = {arXiv},
  primaryclass  = {hep-th},
  slaccitation  = {%%CITATION = ARXIV:1305.2781;%%}
}

@article{Caron-Huot:2016owq,
  author        = {Caron-Huot, Simon and Dixon, Lance J. and McLeod, Andrew
                        and von Hippel, Matt},
  title         = {{Bootstrapping a Five-Loop Amplitude Using Steinmann
                        Relations}},
  journal       = {Phys. Rev. Lett.},
  volume        = {117},
  year          = {2016},
  number        = {24},
  pages         = {241601},
  doi           = {10.1103/PhysRevLett.117.241601},
  eprint        = {1609.00669},
  archiveprefix = {arXiv},
  primaryclass  = {hep-th},
  reportnumber  = {SLAC-PUB-16811},
  slaccitation  = {%%CITATION = ARXIV:1609.00669;%%}
}

@article{Caron-Huot:2018dsv,
  author        = {Caron-Huot, Simon and Dixon, Lance J. and von Hippel,
                        Matt and McLeod, Andrew J. and Papathanasiou, Georgios},
  title         = {{The Double Pentaladder Integral to All Orders}},
  journal       = {JHEP},
  volume        = {07},
  year          = {2018},
  pages         = {170},
  doi           = {10.1007/JHEP07(2018)170},
  eprint        = {1806.01361},
  archiveprefix = {arXiv},
  primaryclass  = {hep-th},
  reportnumber  = {DESY 18-009, SLAC-PUB-17228, DESY-18-041},
  slaccitation  = {%%CITATION = ARXIV:1806.01361;%%}
}

@article{Caron-Huot:2019bsq,
  author        = {Caron-Huot, Simon and Dixon, Lance J. and Dulat, Falko
                        and von Hippel, Matt and McLeod, Andrew J. and
                        Papathanasiou, Georgios},
  title         = {{The Cosmic Galois Group and Extended Steinmann Relations
                        for Planar $\mathcal{N} = 4$ SYM Amplitudes}},
  journal       = {JHEP},
  volume        = {09},
  year          = {2019},
  pages         = {061},
  doi           = {10.1007/JHEP09(2019)061},
  eprint        = {1906.07116},
  archiveprefix = {arXiv},
  primaryclass  = {hep-th},
  reportnumber  = {DESY 19-062, DESY-19-062, HU-EP-19/05, SLAC--PUB--17414},
  slaccitation  = {%%CITATION = ARXIV:1906.07116;%%}
}

@article{Caron-Huot:2019vjl,
  author        = {Caron-Huot, Simon and Dixon, Lance J. and Dulat, Falko
                        and von Hippel, Matt and McLeod, Andrew J. and
                        Papathanasiou, Georgios},
  title         = {{Six-Gluon amplitudes in planar $ \mathcal{N} $ = 4
                        super-Yang-Mills theory at six and seven loops}},
  journal       = {JHEP},
  volume        = {08},
  year          = {2019},
  pages         = {016},
  doi           = {10.1007/JHEP08(2019)016},
  eprint        = {1903.10890},
  archiveprefix = {arXiv},
  primaryclass  = {hep-th},
  reportnumber  = {DESY 19-042, HU-EP-19/04, DESY-19-042, HU-EP-19-04,
                        SLAC-PUB-17413},
  slaccitation  = {%%CITATION = ARXIV:1903.10890;%%}
}

@article{Caron-Huot:2020bkp,
  author        = {Caron-Huot, Simon and Dixon, Lance J. and Drummond, James M. and Dulat, Falko and Foster, Jack and G\"urdo\u{g}an, \"Omer and von Hippel, Matt and McLeod, Andrew J. and Papathanasiou, Georgios},
  title         = {{The Steinmann Cluster Bootstrap for $N$ = 4 Super Yang-Mills Amplitudes}},
  eprint        = {2005.06735},
  archiveprefix = {arXiv},
  primaryclass  = {hep-th},
  reportnumber  = {DESY-20-087},
  doi           = {10.22323/1.376.0003},
  journal       = {PoS},
  volume        = {CORFU2019},
  pages         = {003},
  year          = {2020}
}

@article{CaronHuot:2010ek,
  author        = {Caron-Huot, Simon},
  title         = {{Notes on the scattering amplitude / Wilson loop duality}},
  eprint        = {1010.1167},
  archiveprefix = {arXiv},
  primaryclass  = {hep-th},
  doi           = {10.1007/JHEP07(2011)058},
  journal       = {JHEP},
  volume        = {07},
  pages         = {058},
  year          = {2011}
}

@article{CaronHuot:2011kk,
  author        = {Caron-Huot, Simon and He, Song},
  title         = {{Jumpstarting the All-Loop S-Matrix of Planar N=4 Super
                        Yang-Mills}},
  journal       = {JHEP},
  volume        = {07},
  year          = {2012},
  pages         = {174},
  doi           = {10.1007/JHEP07(2012)174},
  eprint        = {1112.1060},
  archiveprefix = {arXiv},
  primaryclass  = {hep-th},
  slaccitation  = {%%CITATION = ARXIV:1112.1060;%%}
}

@article{Chicherin:2020umh,
    author = "Chicherin, Dimitry and Henn, Johannes M. and Papathanasiou, Georgios",
    title = "{Cluster algebras for Feynman integrals}",
    eprint = "2012.12285",
    archivePrefix = "arXiv",
    primaryClass = "hep-th",
    reportNumber = "DESY-20-204, MPP-2020-216",
    doi = "10.1103/PhysRevLett.126.091603",
    journal = "Phys. Rev. Lett.",
    volume = "126",
    number = "9",
    pages = "091603",
    year = "2021"
}

@article{DelDuca:2011wh,
  author        = {Del Duca, Vittorio and Dixon, Lance J. and Drummond, James M. and Duhr, Claude and Henn, Johannes M. and Smirnov, Vladimir A.},
  title         = {{The one-loop six-dimensional hexagon integral with three massive corners}},
  eprint        = {1105.2011},
  archiveprefix = {arXiv},
  primaryclass  = {hep-th},
  reportnumber  = {HU-EP-11-22, CERN-PH-TH-2011-105, SLAC-PUB-14458, LAPTH-016-11, DCPT-11-42, NSF-KITP-11-072, IPPP-11-21},
  doi           = {10.1103/PhysRevD.84.045017},
  journal       = {Phys. Rev. D},
  volume        = {84},
  pages         = {045017},
  year          = {2011}
}

@article{Dixon:2011pw,
  author        = {Dixon, Lance J. and Drummond, James M. and Henn, Johannes
                        M.},
  title         = {{Bootstrapping the three-loop hexagon}},
  journal       = {JHEP},
  year          = {2011},
  pages         = {023},
  doi           = {10.1007/JHEP11(2011)023},
  eprint        = {1108.4461},
  archiveprefix = {arXiv},
  primaryclass  = {hep-th},
  reportnumber  = {SLAC-PUB-14528, CERN-PH-TH-2011-189, LAPTH-029-11,
                        HU-EP-11-38, NSF-KITP-11-176},
  slaccitation  = {%%CITATION = ARXIV:1108.4461;%%}
}

@article{Dixon:2014iba,
  author        = {Dixon, Lance J. and von Hippel, Matt},
  title         = {{Bootstrapping an NMHV amplitude through three loops}},
  journal       = {JHEP},
  volume        = {10},
  year          = {2014},
  pages         = {065},
  doi           = {10.1007/JHEP10(2014)065},
  eprint        = {1408.1505},
  archiveprefix = {arXiv},
  primaryclass  = {hep-th},
  reportnumber  = {SLAC-PUB-15970},
  slaccitation  = {%%CITATION = ARXIV:1408.1505;%%}
}

@article{Dixon:2014xca,
  author        = {Dixon, Lance J. and Drummond, James M. and Duhr, Claude
                        and von Hippel, Matt and Pennington, Jeffrey},
  title         = {{Bootstrapping six-gluon scattering in planar N=4
                        super-Yang-Mills theory}},
  journal       = {PoS},
  volume        = {LL2014},
  year          = {2014},
  pages         = {077},
  doi           = {10.22323/1.211.0077},
  eprint        = {1407.4724},
  archiveprefix = {arXiv},
  primaryclass  = {hep-th},
  reportnumber  = {SLAC-PUB-16008},
  slaccitation  = {%%CITATION = ARXIV:1407.4724;%%}
}

@article{Dixon:2015iva,
  author        = {Dixon, Lance J. and von Hippel, Matt and McLeod, Andrew
                        J.},
  title         = {{The four-loop six-gluon NMHV ratio function}},
  journal       = {JHEP},
  volume        = {01},
  year          = {2016},
  pages         = {053},
  doi           = {10.1007/JHEP01(2016)053},
  eprint        = {1509.08127},
  archiveprefix = {arXiv},
  primaryclass  = {hep-th},
  reportnumber  = {SLAC-PUB-16352, CALT-2015-049},
  slaccitation  = {%%CITATION = ARXIV:1509.08127;%%}
}

@article{Dixon:2016nkn,
  author        = {Dixon, Lance J. and Drummond, James and Harrington,
                        Thomas and McLeod, Andrew J. and Papathanasiou, Georgios
                        and Spradlin, Marcus},
  title         = {{Heptagons from the Steinmann Cluster Bootstrap}},
  journal       = {JHEP},
  volume        = {02},
  year          = {2017},
  pages         = {137},
  doi           = {10.1007/JHEP02(2017)137},
  eprint        = {1612.08976},
  archiveprefix = {arXiv},
  primaryclass  = {hep-th},
  reportnumber  = {DESY-16-242, Brown-HET-1705, SLAC-PUB-16894},
  slaccitation  = {%%CITATION = ARXIV:1612.08976;%%}
}

@article{Dixon:2020cnr,
  author        = {Dixon, Lance J. and Liu, Yu-Ting},
  title         = {{Lifting Heptagon Symbols to Functions}},
  eprint        = {2007.12966},
  archiveprefix = {arXiv},
  primaryclass  = {hep-th},
  reportnumber  = {SLAC-PUB-17544},
  doi           = {10.1007/JHEP10(2020)031},
  journal       = {JHEP},
  volume        = {10},
  pages         = {031},
  year          = {2020}
}

@article{Dlapa:2021qsl,
    author = "Dlapa, Christoph and Li, Xiaodi and Zhang, Yang",
    title = "{Leading singularities in Baikov representation and Feynman integrals with uniform transcendental weight}",
    eprint = "2103.04638",
    archivePrefix = "arXiv",
    primaryClass = "hep-th",
    reportNumber = "MPP-2021-23, PCFT-21-10, USTC-ICTS-21-10",
    doi = "10.1007/JHEP07(2021)227",
    month = "3",
    year = "2021"
}

@article{Drummond:2006rz,
  author        = {Drummond, J. M. and Henn, J. and Smirnov, V. A. and
                        Sokatchev, E.},
  title         = {{Magic identities for conformal four-point integrals}},
  journal       = {JHEP},
  volume        = {01},
  year          = {2007},
  pages         = {064},
  doi           = {10.1088/1126-6708/2007/01/064},
  eprint        = {hep-th/0607160},
  archiveprefix = {arXiv},
  primaryclass  = {hep-th},
  reportnumber  = {LAPTH-1159-06},
  slaccitation  = {%%CITATION = HEP-TH/0607160;%%}
}

@article{Drummond:2007aua,
  author        = {Drummond, J.M. and Korchemsky, G.P. and Sokatchev, E.},
  title         = {{Conformal properties of four-gluon planar amplitudes and Wilson loops}},
  eprint        = {0707.0243},
  archiveprefix = {arXiv},
  primaryclass  = {hep-th},
  doi           = {10.1016/j.nuclphysb.2007.11.041},
  journal       = {Nucl. Phys. B},
  volume        = {795},
  pages         = {385--408},
  year          = {2008}
}

@article{Drummond:2008vq,
  author        = {Drummond, J. M. and Henn, J. and Korchemsky, G. P. and
                        Sokatchev, E.},
  title         = {{Dual superconformal symmetry of scattering amplitudes in
                        N=4 super-Yang-Mills theory}},
  journal       = {Nucl. Phys.},
  volume        = {B828},
  year          = {2010},
  pages         = {317-374},
  doi           = {10.1016/j.nuclphysb.2009.11.022},
  eprint        = {0807.1095},
  archiveprefix = {arXiv},
  primaryclass  = {hep-th},
  reportnumber  = {LAPTH-1257-08, LPT-ORSAY-08-60},
  slaccitation  = {%%CITATION = ARXIV:0807.1095;%%}
}

@article{Drummond:2009fd,
  author        = {Drummond, James M. and Henn, Johannes M. and Plefka, Jan},
  title         = {{Yangian symmetry of scattering amplitudes in N=4 super
                        Yang-Mills theory}},
  journal       = {JHEP},
  volume        = {05},
  year          = {2009},
  pages         = {046},
  doi           = {10.1088/1126-6708/2009/05/046},
  eprint        = {0902.2987},
  archiveprefix = {arXiv},
  primaryclass  = {hep-th},
  reportnumber  = {HU-EP-09-06, LAPTH-1308-09},
  slaccitation  = {%%CITATION = ARXIV:0902.2987;%%}
}

@article{Drummond:2010cz,
  author        = {Drummond, James M. and Henn, Johannes M. and Trnka, Jaroslav},
  title         = {{New differential equations for on-shell loop integrals}},
  eprint        = {1010.3679},
  archiveprefix = {arXiv},
  primaryclass  = {hep-th},
  reportnumber  = {HU-EP-10-56, CERN-PH-TH-2010-237, LAPTH-042-10},
  doi           = {10.1007/JHEP04(2011)083},
  journal       = {JHEP},
  volume        = {04},
  pages         = {083},
  year          = {2011}
}

@article{Drummond:2014ffa,
  author        = {Drummond, James M. and Papathanasiou, Georgios and
                        Spradlin, Marcus},
  title         = {{A Symbol of Uniqueness: The Cluster Bootstrap for the
                        3-Loop MHV Heptagon}},
  journal       = {JHEP},
  volume        = {03},
  year          = {2015},
  pages         = {072},
  doi           = {10.1007/JHEP03(2015)072},
  eprint        = {1412.3763},
  archiveprefix = {arXiv},
  primaryclass  = {hep-th},
  reportnumber  = {CERN-PH-TH-2014-256, LAPTH-232-14},
  slaccitation  = {%%CITATION = ARXIV:1412.3763;%%}
}

@article{Drummond:2017ssj,
  author        = {Drummond, James and Foster, Jack and G{\"{u}}rdo{\u{g}}an, {\"{O}}mer},
  title         = {{Cluster Adjacency Properties of Scattering Amplitudes in
                        $N=4$ Supersymmetric Yang-Mills Theory}},
  journal       = {Phys. Rev. Lett.},
  volume        = {120},
  year          = {2018},
  number        = {16},
  pages         = {161601},
  doi           = {10.1103/PhysRevLett.120.161601},
  eprint        = {1710.10953},
  archiveprefix = {arXiv},
  primaryclass  = {hep-th},
  slaccitation  = {%%CITATION = ARXIV:1710.10953;%%}
}

@article{Drummond:2018caf,
  author        = {Drummond, James and Foster, Jack and G{\"{u}}rdo{\u{g}}an, {\"{O}}mer
                        and Papathanasiou, Georgios},
  title         = {{Cluster adjacency and the four-loop NMHV heptagon}},
  journal       = {JHEP},
  volume        = {03},
  year          = {2019},
  pages         = {087},
  doi           = {10.1007/JHEP03(2019)087},
  eprint        = {1812.04640},
  archiveprefix = {arXiv},
  primaryclass  = {hep-th},
  reportnumber  = {DESY-18-214},
  slaccitation  = {%%CITATION = ARXIV:1812.04640;%%}
}

@article{Drummond:2018dfd,
  author        = {Drummond, James and Foster, Jack and G{\"{u}}rdo{\u{g}}an, {\"{O}}mer},
  title         = {{Cluster adjacency beyond MHV}},
  journal       = {JHEP},
  volume        = {03},
  year          = {2019},
  pages         = {086},
  doi           = {10.1007/JHEP03(2019)086},
  eprint        = {1810.08149},
  archiveprefix = {arXiv},
  primaryclass  = {hep-th},
  slaccitation  = {%%CITATION = ARXIV:1810.08149;%%}
}

@article{Drummond:2019cxm,
    author = {Drummond, James and Foster, Jack and G\"urdogan, \"Omer and Kalousios, Chrysostomos},
    title = "{Algebraic singularities of scattering amplitudes from tropical geometry}",
    eprint = "1912.08217",
    archivePrefix = "arXiv",
    primaryClass = "hep-th",
    doi = "10.1007/JHEP04(2021)002",
    journal = "JHEP",
    volume = "04",
    pages = "002",
    year = "2021"
}

@article{Drummond:2019qjk,
    author = {Drummond, James and Foster, Jack and G\"urdogan, \"Omer and Kalousios, Chrysostomos},
    title = "{Tropical Grassmannians, cluster algebras and scattering amplitudes}",
    eprint = "1907.01053",
    archivePrefix = "arXiv",
    primaryClass = "hep-th",
    doi = "10.1007/JHEP04(2020)146",
    journal = "JHEP",
    volume = "04",
    pages = "146",
    year = "2020"
}

@article{Duhr:2011zq,
  author        = {Duhr, Claude and Gangl, Herbert and Rhodes, John R.},
  title         = {{From polygons and symbols to polylogarithmic functions}},
  journal       = {JHEP},
  volume        = {10},
  year          = {2012},
  pages         = {075},
  doi           = {10.1007/JHEP10(2012)075},
  eprint        = {1110.0458},
  archiveprefix = {arXiv},
  primaryclass  = {math-ph},
  reportnumber  = {IPPP-11-56, DCPT-11-112},
  slaccitation  = {%%CITATION = ARXIV:1110.0458;%%}
}

@article{Elvang:2014fja,
    author = "Elvang, Henriette and Huang, Yu-tin and Keeler, Cynthia and Lam, Thomas and Olson, Timothy M. and Roland, Samuel B. and Speyer, David E.",
    title = "{Grassmannians for scattering amplitudes in 4d $\mathcal{N}=4$ SYM and 3d ABJM}",
    eprint = "1410.0621",
    archivePrefix = "arXiv",
    primaryClass = "hep-th",
    reportNumber = "MCTP-14-36",
    doi = "10.1007/JHEP12(2014)181",
    journal = "JHEP",
    volume = "12",
    pages = "181",
    year = "2014"
}

@article{fomin2002cluster,
  title   = {Cluster algebras I: foundations},
  author  = {Fomin, Sergey and Zelevinsky, Andrei},
  journal = {Journal of the American Mathematical Society},
  volume  = {15},
  number  = {2},
  pages   = {497--529},
  year    = {2002}
}

@article{fomin2003cluster,
  title     = {Cluster algebras II: Finite type classification},
  author    = {Fomin, Sergey and Zelevinsky, Andrei},
  journal   = {Inventiones mathematicae},
  volume    = {154},
  number    = {1},
  pages     = {63--121},
  year      = {2003},
  publisher = {Springer}
}

@article{fomin2007cluster,
  title     = {Cluster algebras IV: coefficients},
  author    = {Fomin, Sergey and Zelevinsky, Andrei},
  journal   = {Compositio Mathematica},
  volume    = {143},
  number    = {1},
  pages     = {112--164},
  year      = {2007},
  publisher = {London Mathematical Society}
}

@article{Golden:2013xva,
  author        = {Golden, John and Goncharov, Alexander B. and Spradlin,
                        Marcus and Vergu, Cristian and Volovich, Anastasia},
  title         = {{Motivic Amplitudes and Cluster Coordinates}},
  journal       = {JHEP},
  volume        = {01},
  year          = {2014},
  pages         = {091},
  doi           = {10.1007/JHEP01(2014)091},
  eprint        = {1305.1617},
  archiveprefix = {arXiv},
  primaryclass  = {hep-th},
  slaccitation  = {%%CITATION = ARXIV:1305.1617;%%}
}

@article{Golden:2014xqa,
  author        = {Golden, John and Paulos, Miguel F. and Spradlin, Marcus
                        and Volovich, Anastasia},
  title         = {{Cluster Polylogarithms for Scattering Amplitudes}},
  journal       = {J. Phys.},
  volume        = {A47},
  year          = {2014},
  number        = {47},
  pages         = {474005},
  doi           = {10.1088/1751-8113/47/47/474005},
  eprint        = {1401.6446},
  archiveprefix = {arXiv},
  primaryclass  = {hep-th},
  reportnumber  = {BROWN-HET-1654},
  slaccitation  = {%%CITATION = ARXIV:1401.6446;%%}
}

@article{Goncharov:2010jf,
  author        = {Goncharov, Alexander B. and Spradlin, Marcus and Vergu,
                        C. and Volovich, Anastasia},
  title         = {{Classical Polylogarithms for Amplitudes and Wilson
                        Loops}},
  journal       = {Phys. Rev. Lett.},
  volume        = {105},
  year          = {2010},
  pages         = {151605},
  doi           = {10.1103/PhysRevLett.105.151605},
  eprint        = {1006.5703},
  archiveprefix = {arXiv},
  primaryclass  = {hep-th},
  reportnumber  = {BROWN-HET-1602},
  slaccitation  = {%%CITATION = ARXIV:1006.5703;%%}
}

@article{He:2020lcu,
    author = "He, Song and Li, Zhenjie and Yang, Qinglin and Zhang, Chi",
    title = "{Feynman Integrals and Scattering Amplitudes from Wilson Loops}",
    eprint = "2012.15042",
    archivePrefix = "arXiv",
    primaryClass = "hep-th",
    doi = "10.1103/PhysRevLett.126.231601",
    journal = "Phys. Rev. Lett.",
    volume = "126",
    pages = "231601",
    year = "2021"
}

@article{He:2020uhb,
  author        = {He, Song and Li, Zhenjie},
  title         = {{A Note on Letters of Yangian Invariants}},
  eprint        = {2007.01574},
  archiveprefix = {arXiv},
  primaryclass  = {hep-th},
  doi           = {10.1007/JHEP02(2021)155},
  journal       = {JHEP},
  volume        = {02},
  pages         = {155},
  year          = {2021}
}

@article{He:2020uxy,
    author = "He, Song and Li, Zhenjie and Tang, Yichao and Yang, Qinglin",
    title = "{The Wilson-loop $d$ log representation for Feynman integrals}",
    eprint = "2012.13094",
    archivePrefix = "arXiv",
    primaryClass = "hep-th",
    doi = "10.1007/JHEP05(2021)052",
    journal = "JHEP",
    volume = "05",
    pages = "052",
    year = "2021"
}

@article{He:2020vob,
    author = "He, Song and Li, Zhenjie and Zhang, Chi",
    title = "{The symbol and alphabet of two-loop NMHV amplitudes from $\bar{Q}$ equations}",
    eprint = "2009.11471",
    archivePrefix = "arXiv",
    primaryClass = "hep-th",
    doi = "10.1007/JHEP03(2021)278",
    journal = "JHEP",
    volume = "03",
    pages = "278",
    year = "2021"
}

@article{Henke:2019hve,
  author        = {Henke, Niklas and Papathanasiou, Georgios},
  title         = {{How tropical are seven- and eight-particle amplitudes?}},
  eprint        = {1912.08254},
  archiveprefix = {arXiv},
  primaryclass  = {hep-th},
  reportnumber  = {DESY-19-229},
  doi           = {10.1007/JHEP08(2020)005},
  journal       = {JHEP},
  volume        = {08},
  pages         = {005},
  year          = {2020}
}

@article{Henn:2013pwa,
  author        = {Henn, Johannes M.},
  title         = {{Multiloop integrals in dimensional regularization made simple}},
  eprint        = {1304.1806},
  archiveprefix = {arXiv},
  primaryclass  = {hep-th},
  doi           = {10.1103/PhysRevLett.110.251601},
  journal       = {Phys. Rev. Lett.},
  volume        = {110},
  pages         = {251601},
  year          = {2013}
}

@article{Henn:2014qga,
  author        = {Henn, Johannes M.},
  title         = {{Lectures on differential equations for Feynman
                        integrals}},
  journal       = {J. Phys.},
  volume        = {A48},
  year          = {2015},
  pages         = {153001},
  doi           = {10.1088/1751-8113/48/15/153001},
  eprint        = {1412.2296},
  archiveprefix = {arXiv},
  primaryclass  = {hep-ph},
  slaccitation  = {%%CITATION = ARXIV:1412.2296;%%}
}

@article{Henn:2018cdp,
  author        = {Henn, Johannes and Herrmann, Enrico and Parra-Martinez,
                        Julio},
  title         = {{Bootstrapping two-loop Feynman integrals for planar $
                        \mathcal{N}=4 $ sYM}},
  journal       = {JHEP},
  volume        = {10},
  year          = {2018},
  pages         = {059},
  doi           = {10.1007/JHEP10(2018)059},
  eprint        = {1806.06072},
  archiveprefix = {arXiv},
  primaryclass  = {hep-th},
  reportnumber  = {MPP-2018-135, MITP/18-049, MITP-18-049},
  slaccitation  = {%%CITATION = ARXIV:1806.06072;%%}
}

@article{Herrmann:2019upk,
  author        = {Herrmann, Enrico and Parra-Martinez, Julio},
  title         = {{Logarithmic forms and differential equations for Feynman integrals}},
  eprint        = {1909.04777},
  archiveprefix = {arXiv},
  primaryclass  = {hep-th},
  doi           = {10.1007/JHEP02(2020)099},
  journal       = {JHEP},
  volume        = {02},
  pages         = {099},
  year          = {2020}
}

@article{Hodges:2009hk,
  author        = {Hodges, Andrew},
  title         = {{Eliminating spurious poles from gauge-theoretic
                        amplitudes}},
  journal       = {JHEP},
  volume        = {05},
  year          = {2013},
  pages         = {135},
  doi           = {10.1007/JHEP05(2013)135},
  eprint        = {0905.1473},
  archiveprefix = {arXiv},
  primaryclass  = {hep-th},
  slaccitation  = {%%CITATION = ARXIV:0905.1473;%%}
}

@article{Mago:2020kmp,
  author        = {Mago, Jorge and Schreiber, Anders and Spradlin, Marcus and Volovich, Anastasia},
  title         = {{Symbol alphabets from plabic graphs}},
  eprint        = {2007.00646},
  archiveprefix = {arXiv},
  primaryclass  = {hep-th},
  doi           = {10.1007/JHEP10(2020)128},
  journal       = {JHEP},
  volume        = {10},
  pages         = {128},
  year          = {2020}
}

@article{Mago:2020nuv,
    author = "Mago, J. and Schreiber, A. and Spradlin, M. and Yelleshpur Srikant, A. and Volovich, A.",
    title = "{Symbol alphabets from plabic graphs II: rational letters}",
    eprint = "2012.15812",
    archivePrefix = "arXiv",
    primaryClass = "hep-th",
    doi = "10.1007/JHEP04(2021)056",
    journal = "JHEP",
    volume = "04",
    pages = "056",
    year = "2021"
}

@article{Mago:2021luw,
    author = "Mago, Jorge and Schreiber, Anders and Spradlin, Marcus and Yelleshpur Srikant, Akshay and Volovich, Anastasia",
    title = "{Symbol Alphabets from Plabic Graphs III: n=9}",
    eprint = "2106.01406",
    archivePrefix = "arXiv",
    primaryClass = "hep-th",
    month = "6",
    year = "2021"
}

@article{speyer2005tropical,
  title     = {The tropical totally positive Grassmannian},
  author    = {Speyer, David and Williams, Lauren},
  journal   = {Journal of Algebraic Combinatorics},
  volume    = {22},
  number    = {2},
  pages     = {189--210},
  year      = {2005},
  publisher = {Springer}
}

@article{Spradlin:2011wp,
  author        = {Spradlin, Marcus and Volovich, Anastasia},
  title         = {{Symbols of One-Loop Integrals From Mixed Tate Motives}},
  eprint        = {1105.2024},
  archiveprefix = {arXiv},
  primaryclass  = {hep-th},
  reportnumber  = {BROWN-HET-1612, NSF-KITP-11-076},
  doi           = {10.1007/JHEP11(2011)084},
  journal       = {JHEP},
  volume        = {11},
  pages         = {084},
  year          = {2011}
}

@article{Zhang:2019vnm,
  author        = {He, Song and Li, Zhenjie and Zhang, Chi},
  title         = {{Two-loop Octagons, Algebraic Letters and $\bar{Q}$ Equations}},
  eprint        = {1911.01290},
  archiveprefix = {arXiv},
  primaryclass  = {hep-th},
  doi           = {10.1103/PhysRevD.101.061701},
  journal       = {Phys. Rev. D},
  volume        = {101},
  number        = {6},
  pages         = {061701},
  year          = {2020}
}

@article{He:2021esx,
    author = "He, Song and Li, Zhenjie and Yang, Qinglin",
    title = "{Notes on cluster algebras and some all-loop Feynman integrals}",
    eprint = "2103.02796",
    archivePrefix = "arXiv",
    primaryClass = "hep-th",
    doi = "10.1007/JHEP06(2021)119",
    journal = "JHEP",
    volume = "06",
    pages = "119",
    year = "2021"
}

@article{He:2021fwf,
    author = "He, Song and Li, Zhenjie and Tang, Yichao and Yang, Qinglin",
    title = "{Bootstrapping octagons in reduced kinematics from $A_2$ cluster algebras}",
    eprint = "2106.03709",
    archivePrefix = "arXiv",
    primaryClass = "hep-th",
    month = "6",
    year = "2021"
}

@article{Henke:2021avn,
    author = "Henke, Niklas and Papathanasiou, Georgios",
    title = "{Singularities of eight- and nine-particle amplitudes from cluster algebras and tropical geometry}",
    eprint = "2106.01392",
    archivePrefix = "arXiv",
    primaryClass = "hep-th",
    month = "6",
    year = "2021"
}

@article{Ren:2021ztg,
    author = "Ren, Lecheng and Spradlin, Marcus and Volovich, Anastasia",
    title = "{Symbol Alphabets from Tensor Diagrams}",
    eprint = "2106.01405",
    archivePrefix = "arXiv",
    primaryClass = "hep-th",
    month = "6",
    year = "2021"
}

@article{speyer2004tropical,
  title={The tropical grassmannian},
  author={Speyer, David and Sturmfels, Bernd},
  year={2004},
  publisher={Walter de Gruyter GmbH \& Co. KG Berlin, Germany}
}

@article{He:2021non,
    author = "He, Song and Li, Zhenjie and Yang, Qinglin",
    title = "{Truncated cluster algebras and Feynman integrals with algebraic letters}",
    eprint = "2106.09314",
    archivePrefix = "arXiv",
    primaryClass = "hep-th",
    month = "6",
    year = "2021"
}

\end{document}